\documentclass[twocolumn, tighten, trackchanges]{aastex63}
\usepackage{amsmath,amsopn,amsxtra,txfonts}
\usepackage{comment} 
\usepackage{quotes}

\newcommand{\ha}{H$\alpha$}
\newcommand{\kms}{ \ifmmode{\rm km\thinspace s^{-1}}\else km\thinspace s$^{-1}$\fi}
\newcommand{\kpc}{\ensuremath{\, \mathrm{kpc}}}
\newcommand{\cm}{\ensuremath{ \, \mathrm{cm}}}
\newcommand{\s}{\ensuremath{ \, \mathrm{s}}}
\newcommand{\sr}{\ensuremath{ \, \mathrm{sr}}}

\newcommand{\dg}{\ifmmode{^{\circ}}\else $^{\circ}$\fi}

\newcommand{\hi}{H\textsc{~i}}
\newcommand{\vlsr}{\ifmmode{{\rm v}_{\rm{LSR}}}\else ${\rm v}_{\rm{LSR}}$\fi}
\newcommand{\iha}{\ifmmode{I_{\rm{H}\alpha}} \else $I_{\rm H \alpha}$\fi}

\newcommand{\mnhi}{N_{{\rm H}\textsc{~i}}}
\newcommand{\avN}{$\langle N_{H\textsc{~i}} \rangle$}

\newcommand{\vgeo}{\ifmmode{{\rm v}_{\mathrm{GEO}}} \else ${\rm v}_{\mathrm{GEO}}$\fi}

\defcitealias{Putman2003}{PBV03}
\defcitealias{Mcclure-Griffiths2008}{MSL08}
\defcitealias{Zhang2019}{ZCB19}
\defcitealias{PriceWhelan2019}{PAN19}
\defcitealias{Smoker2011}{SFK11}
\defcitealias{Venzmer2012}{VKK12}
\defcitealias{Pardy2018}{PDF18}

\begin{document}
\title{\ha\ Distances to the Leading Arm of the Magellanic Stream}
	\shorttitle{\ha\ Distances to the Leading Arm of the Magellanic Stream}
	\shortauthors{Antwi-Danso, Barger, and Haffner} 
 
\author[0000-0002-0243-6575]{Jacqueline Antwi-Danso}
\affiliation{George P. and Cynthia Woods Mitchell Institute for Fundamental Physics and Astronomy, Texas A\&M University, College Station, TX 78743, USA}
\affiliation{Department of Physics and Astronomy, Texas A\&M University, 4242 TAMU, College Station, TX 78743, USA}
\affiliation{Department of Physics \& Astronomy, Texas Christian University, Fort Worth, TX 76129, USA}
\author[0000-0001-5817-0932]{Kathleen A. Barger}
\affiliation{Department of Physics \& Astronomy, Texas Christian University, Fort Worth, TX 76129, USA}
\affiliation{Department of Physics, University of Notre Dame, Notre Dame, IN 46556 , USA}
\author[0000-0002-9947-6396]{L. Matthew Haffner}
\affiliation{Embry-Riddle Aeronautical University, Daytona Beach, FL 32114, USA}
\affiliation{Space Science Institute, Boulder, CO 80301, USA}
\affiliation{Department of Astronomy, University of Wisconsin-Madison, Madison, WI 53706, USA}

\correspondingauthor{Jacqueline Antwi-Danso}
\email{jadanso@tamu.edu}

\received{April 30, 2019}
\revised{December 18, 2019}
\accepted{January 21, 2020}
\published{March 10, 2020}
\submitjournal{\apj}

\begin{abstract}
The Leading Arm is a tidal feature that is in front of the Magellanic Clouds on their orbit through the Galaxy's halo. Many physical properties of the Leading Arm, such as its mass and size, are poorly constrained because it has few distance measurements. While \ha\ measurements have been used to estimate the distances to halo clouds, many studies have been unsuccessful in detecting \ha\ from the Leading Arm. In this study, we explore a group of \hi\ clouds which lie $75\dg - 90\arcdeg$ from the Magellanic Clouds. Through ultraviolet and 21-cm radio spectroscopy, this region, dubbed the Leading Arm Extension, was found to have chemical and kinematic similarities to the Leading Arm. Using the Wisconsin \ha\ Mapper, we detect \ha\ emission in four out of seven of our targets. Assuming that this region is predominantly photoionized, we use a radiation model that incorporates the contributions of the Galaxy, Magellanic Clouds, and the extragalactic background at $\rm z = 0$ to derive a heliocentric distance of $d_{\odot}\ge13.4~\kpc$. We also use this model to rederive \ha\ distances of $d_{\odot} \geq 5.0$ kpc and $d_{\odot} \geq 22.9~\kpc$ to two clouds in the literature that might also be associated with the Leading Arm. Using these new measurements, and others in the literature, we provide a general trend of the variation of Leading Arm heliocentric distance as a function of Magellanic Stream longitude, and explore its implications for the origin and closest point of approach of the Leading Arm.  
\end{abstract}

\keywords{galaxies: Magellanic Clouds -- galaxies: dwarf -- Galaxy: evolution -- Galaxy: halo -- ISM: individual (Leading Arm)}

\section{Introduction}\label{section:introduction}
The Milky Way's circumgalactic medium contains numerous high-velocity clouds (HVCs). These have various origins, including aggregates of neutral and ionized hydrogen from satellite galaxies, gas cooling out of the intergalactic medium, and Galactic fountains. As suggested by stellar chemical evolution models of the Milky Way (MW) (e.g., \citealt{Chiappini2008}), some HVCs serve as a source of low metallicity gas on which the Galaxy sustains its star formation. Therefore, a key component of understanding the relationship between baryonic feedback processes and the Galaxy's evolution relies on accurate measurements of HVC physical properties. HVC distances are especially important for estimating their basic physical properties, because many of them scale directly with distance ($d$). For example, their mass, size, pressure, and density scale as ${\rm M}_{\rm cloud}\propto d^{2}$, $D_{\rm cloud} \propto d$, ${\rm P}_{\rm cloud}\propto d^{-1}$, and $n_{\rm cloud}\propto d^{-1}$, respectively.

Spanning about 11,000 square degrees in both ionized and neutral gas \citep{Fox2014}, the gaseous streams of the interacting Magellanic Clouds (MCs) --- known as the Magellanic System (MSys, see \citealt{DOnghia2016} for a review) --- are the largest collection of HVCs surrounding the MW. The MSys extends over $\rm 200\dg$ across the sky, and is comprised of the trailing Stream, the Bridge that connects the Large and Small Magellanic Clouds (LMC and SMC), and the Leading Arm (LA). As its name implies, the LA is the gaseous counterpart to the Stream that leads the MCs on their orbit through the Galactic potential.

\begin{figure*}
\begin{center}
\includegraphics[scale=0.45,angle=0]{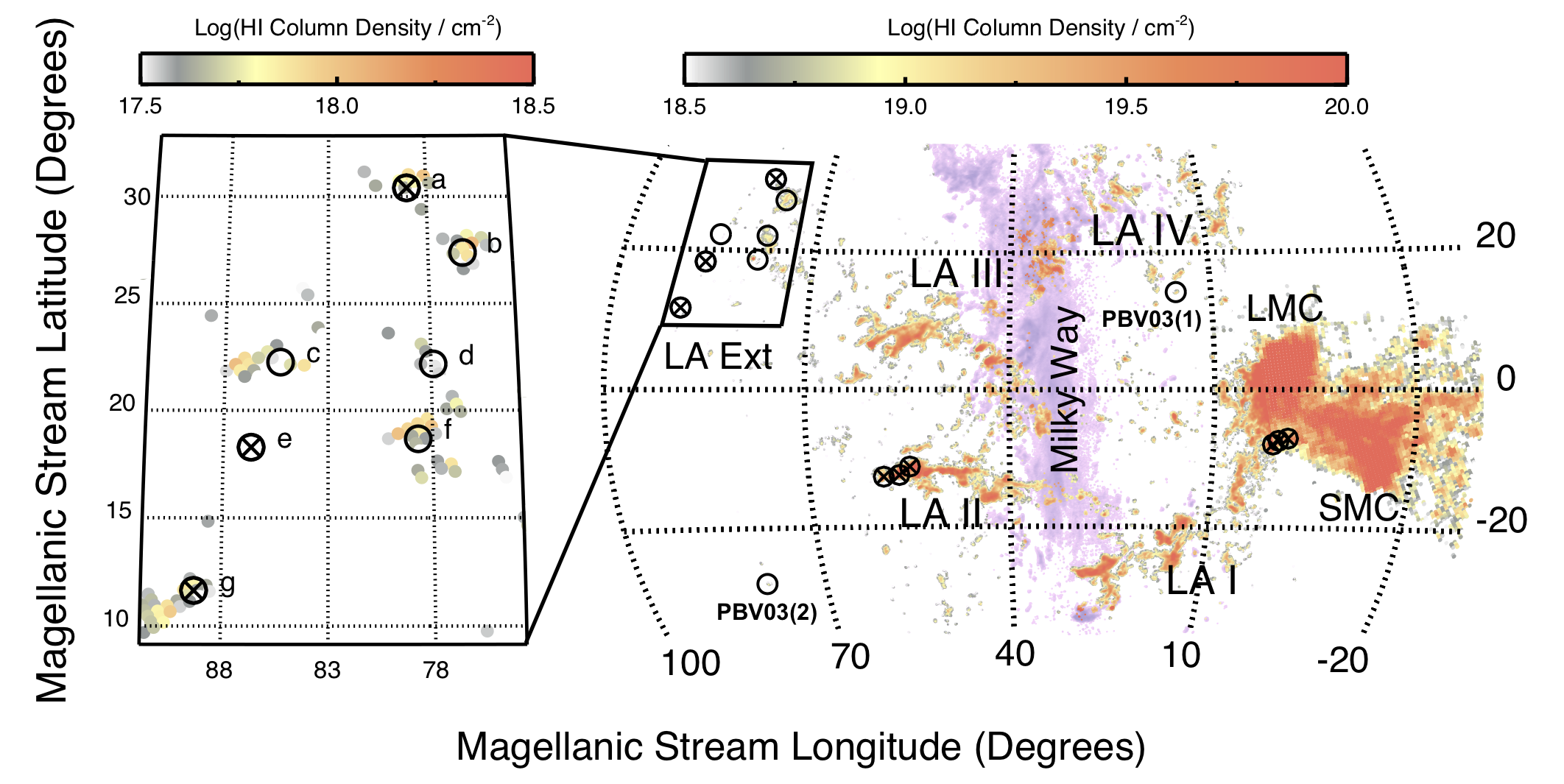}
\end{center}
\figcaption{An \hi\ emission map of the Leading Arm and MCs in Magellanic Stream coordinates. The locations of our \ha\ observations are labeled~(a) through~(g) in the zoomed-in region. This is shown against a backdrop of the LAB \hi\ survey column densities from the Gaussian decompositions of \citet{Nidever2008}. The  open circles indicate detections and cross-filled circles non-detections. The \ha\ detections labelled \citetalias{Putman2003}(1) and \citetalias{Putman2003}(2) are the \citep{Putman2003} detections toward CHVC+266.0-18.7+336 and HVC+310.5+44.2+187, respectively. The non-detections in LA~II and the outskirts of the LMC are from \citet{McClure-Griffiths2010}.  The global \hi\ distribution map on the right hand side shows Galactic All-Sky Survey (GASS) observations that have been integrated over $+150~\lesssim\vlsr\lesssim+350~\kms$ \citep{McClure-Griffiths2009}. The \hi\ emission from the MW is shown in purple to distinguish it from the Magellanic emission. 
\label{fig:all}}
\end{figure*}

The LA covers approximately $\rm 60\dg\times80\dg$ on the sky \citep{Fox2018} and is separated into four main complexes named LA~I--IV. It has a complicated velocity structure with localstandard of rest (LSR) velocities \footnote{Throughout this study, we use the kinematic definition of the LSR, where the solar motion is $20~\kms$ toward $(\rm \alpha,\delta)_{\rm J2000}=(18^h 3^m 50.29^s, 30\arcdeg00\arcmin16\farcs8)$.}  spanning $+70\lesssim\vlsr\lesssim+500~\kms$ \citep{Wakker1991,Bruns2005,Richter2017a,Fox2018}. This fragmented structure has been studied extensively in \hi~21-cm emission (\citealt{Putman1998, Putman2002,Putman2003, Bruns2005, Mcclure-Griffiths2008, McClure-Griffiths2010, Nidever2008, Nidever2010, Venzmer2012, For2013, For2016}, among others) and UV absorption \citep{1994ApJ...426..563L, 1998AJ....115..162L,2001AJ....121..992S, Fox2014, Fox2018, Richter2018}. These four complexes are accompanied by less-studied, smaller clouds that are moving at similar velocities along the same orbital path (Figure~\ref{fig:all}). Because of its complicated structure, determining the total mass of the LA requires distance estimates at multiple locations and along the full extent of the LA. These distances provide important observational constraints for modeling the interactions of the MCs and their passage through the MW's halo (\citealt{Besla2007, Besla2010, Besla2012a}, \citealt{Diaz2011}; \citealt{Guglielmo2014}; \citealt{Pardy2018}). 

One can determine HVC distances using direct or indirect methods. Previous LA studies have employed both. The former approach involves finding the distance to stars formed in-situ, or bracketing the distance to the HVC by measuring absorption (or the lack of) at the velocities of the HVC in the spectra of stars whose distances are known (using them as background targets). \citet{Smoker2011} found a lack of interstellar absorption at high velocities toward the star HD~86248, which gives a lower distance limit to LA~I (formerly called Complex~EP) of $d_{\odot}\ge 5.9~\kpc$. Also in the direction of LA~I, \citet{Casetti-Dinescu2014} identified a group of young stars at $12\le d_{\odot} \le 21~\kpc$. They were originally thought to have formed within the LA, and therefore their distance would have been a direct distance measurement to the LA. However, \citet{Zhang2019} recently used {\em Gaia} proper motion measurements of these stars to show that they are moving with the disk of the Milky Way and are therefore not associated with the LA. On the other hand, \citet{Fox2018} used one of these stars, CD14-A05, as a background target and found absorption at LA velocities in its spectrum. Using {\em Gaia} DR2 parallaxes and high-resolution MIKE spectra, \citet{Zhang2019} found that this star is at $d_{\odot} = 22.0\pm3.0~\kpc$, which places a 1$\sigma$ upper distance limit to the absorbing material of LA~I at $d_{\odot} \lesssim 25~\kpc$. More recently, \citet{PriceWhelan2019} used {\em Gaia} astrometry and DECam optical images to isolate a low metallicity ($ \rm [Fe/H] \approx -1.1$) young stellar group in the direction of LA~II. This chemical composition is  consistent with chemical abundances found by \citet{Fox2018} and \citet{Richter2018} for the LA. \cite{PriceWhelan2019} found a distance of $ d_{\odot} = 28.9 \pm 0.1~\kpc$ to the stellar association in LA~II.

One can also use models constrained by observations to indirectly estimate distances to HVCs. Using \hi\ observations taken with the Australia Telescope Compact Array, \citet{Mcclure-Griffiths2008}  showed evidence of an interaction between the LA~I cloud HVC~306-2+230 and the Galactic disk. The interaction with neutral gas at Galactic velocities places the high latitude tip of LA~I at a heliocentric distance of $d_{\odot}=21 \pm 4.2~\kpc$. 
By estimating the minimum travel time for the LA~I complex from the LMC assuming the LA is viewed nearly face-on, \citet{Venzmer2012} estimated a distance of $d_{\odot}\approx 23.5~\kpc$ to LA~I, which is consistent with \cite{Mcclure-Griffiths2008}. Using this same technique, \citep{Venzmer2012} estimated that LA~IV lies at $d_{\odot}\approx 74~\kpc$.

The \ha\ recombination line can be used to estimate HVC distances using a three-dimensional model of the Galactic Lyman continuum flux ($\phi_{\rm LyC}$) if one assumes that photoionization is the dominant source of ionization. \citep{Putman2003} detected \ha\ toward CHVC266.0-18.7+336 in LA~IV and HVC310.5+44.2+187, which is located on the outskirts of LA II (Figure \ref{fig:all}). They used these \ha\ emission-line observations and the \citet{BH1999,BH2001} model of the Galactic ionizing radiation to derive distances of $1.2\le d_{\odot}\le 6.1~\kpc$ and $0.4\le d_{\odot}\le27.5~\kpc$, respectively to these HVCs. \citet{Putman2003} is the only study to do this successfully till date, as the LA is notoriously faint in \ha.\ Rather, \ha\ non-detections have been reported along the LA, even using deep \ha\ spectroscopic surveys (\citealt{Putman2003}; \citealt{McClure-Griffiths2010}).

In this paper, we present seven new targeted \ha\ pointings toward six small clouds positioned $75\dg - 90\arcdeg$ from the LMC (Figure~\ref{fig:all}). \citet{Fox2018} explored the O/H abundances of these clouds using {\em Hubble Space Telescope}/Cosmic Origins Spectrograph ({\em HST}/COS) absorption-line spectroscopy toward background quasars and found that the clouds are likely members of the LA based on their composition, position, and velocity. Since these clouds are leading the LA~III complex by roughly $15\dg-30\dg$, they might represent a diffuse leading edge of the complex. \citet{Fox2018} named this region the Leading Arm Extension (LA~Ext) because it comprises smaller \hi\ clouds located to the north of the main LA complexes. We compare our observations with the {\em HST}/COS absorption-line observations of \citet{Fox2014, Fox2018} to investigate the ionization conditions of this gas in Section~\ref{section:compare}. In Section~\ref{section:dist}, we use models of the ionizing radiation from the MW \citep{Fox2014}, Magellanic Clouds \citep{Barger2013}, and EGB \citep{Weymann2001} to constrain the distances of these clouds and rederive the distances to CHVC+266.0-18.7+336 and HVC+310.5+44.2+187, which were previously detected in \ha\ by \citep{Putman2003} and \citet{McClure-Griffiths2010}, respectively. Finally in Section \ref{section:disc}, we explore the implications of the trend in distance to the Leading Arm on issues that are still in debate, namely the origin of the LA, and its point of closest approach.

\begin{deluxetable*}{crccrcccccrccccrccccc}
\tabletypesize{\scriptsize}
\tablecaption{Summary of Results for LA~Ext Cloudlets and detected \ha\ emission in the LA region from other studies \label{table:intensities}}
\tablewidth{0pt}
\tablehead{
\colhead{} &&\multicolumn{2}{c}{Coordinates} && \multicolumn{5}{c}{\ha} & & \multicolumn{4}{c}{H\textsc{~i}\tablenotemark{d}} & \colhead{} & \colhead{Distance}\\
\cline{3-4}\cline{6-10}\cline{12-15}\cline{17-17}
\colhead{ID} && \colhead{$l,~ b$} & \colhead{$l_{MS},~ b_{MS}$\tablenotemark{a}}&& \colhead{\iha\tablenotemark{b}} & \colhead{$f_{\rm ext,~corr}$\tablenotemark{c}} & \colhead{$\vlsr$} & \colhead{FWHM}  & \colhead{$\tilde{\chi}^{2}_{\rm min}$} & & 
                         \colhead{$\log{\left(N_{\rm H\textsc{~i}}/{\rm cm}^{-2}\right)}$}&\colhead{$\vlsr$} & \colhead{FWHM} & \colhead{$\tilde{\chi}^{2}_{\rm min}$} & & \colhead{$d_{\odot}$} \\
\colhead{} && \colhead{(\dg)} &\colhead{(\dg)}&& \colhead{(mR)} & \colhead{(\%)} & \colhead{($\kms$)} & \colhead{($\kms$)} &  & & \colhead{} & \colhead{($\kms$)} & \colhead{($\kms$)} &&& \colhead{(kpc)} 
}
\startdata
a &&$234.1$, $33.5$&$79.0$, $30.5$&& $<47$\tablenotemark{e} &  $5.6$ & \nodata & \nodata & \nodata &  & $19.24\pm0.01$ & $126.1\pm0.9$ & $40.6\pm0.8$ & $1.1$ && \nodata \\ 
b &&$238.7$, $33.1$ & $76.3$, $27.4$ && $46\pm6 ^{+23}_{-21}$ & $6.4$ & $145\pm7$ & $61\pm5$ & $1.3$ && $19.15\pm0.01$&$ 154.7\pm 0.7$&$30.8\pm 0.7 $&$ 1.1 $ && $10.4^{+0.8~(+6.5)}_{-0.7~(-2.9)}$ \\ 
c &&$238.5$, $42.8$&$85.3$, $22.3$ &&$40\pm10 ^{+15}_{-20}$ & $4.7$ & $73\pm7$ & $47\pm7$ & $2.1$ && $18.88\pm0.03$ & $64.1\pm0.8$ & $22.6\pm 1.4$ & $1.2$ && $17.5^{+4.9~(+8.1)}_{-3.5~(-5.3)}$  \\   
&&$238.5$, $42.8$ & $85.3$, $22.3$&& $31\pm7^{+15}_{-18}$ & $4.7$ & $112  \pm7$ & $45\pm13$ & $1.9$ && $19.01\pm0.02$ & $111.2\pm1.0$ & $28.2\pm   1.2$&$ 1.1 $ && $21.3^{+4.3~(+16.7)}_{-3.0~(-7.3)}$  \\ 
d &&$243.3$, $37.0$ &$77.7$, $22.2$ && $39\pm3^{+10}_{-9}$ & $5.4$ & $ 153\pm6$ & $45\pm1$ & $1.2$ && $18.96\pm0.03 $&$ 166.1\pm 0.9$&$36.7\pm 1.3$&$ 1.0 $ && $15.1^{+0.6~(+3.7)}_{-0.5~(-2.2)}$ \\ 
e &&$242.2$, $46.1$&$86.7$, $18.3$&& $<33$\tablenotemark{e} & $8.5$ & \nodata & \nodata & \nodata && $<18.9$\tablenotemark{e} & \nodata & \nodata & \nodata && \nodata \\ 
f &&$246.8$, $39.3$ & $78.7$, $18.7$&&$46\pm4^{+15}_{-14}$ & $7.4$ & $121\pm 7$ & $61\pm3$ & $1.0$ && $19.25\pm0.01$ & $126.6\pm0.7$ & $42.1\pm0.8$ & $1.0$ && $13.7^{+0.8~(+4.9)}_{-0.5~(-2.5)}$ \\ 
g &&$248.9$, $51.6$&$89.3$, $11.6$&& $<28$\tablenotemark{e} & $8.4$ & \nodata & \nodata & \nodata && $19.13\pm0.01 $&$97.2\pm 0.5$&$25.1\pm0.7$ & $1.1$ && \nodata \\ 
\hline 
\citetalias{Putman2003}(1)\tablenotemark{f} &&$266.0$, $-18.7$&$15.3$, $14.2$&&$136-187\tablenotemark{g}$ & $18.0$ & $336$& \nodata & \nodata && $19.15$ &$336$ &$31$ & \nodata &&$5.0-5.6$  \\
\citetalias{Putman2003}(2)\tablenotemark{f} &&$310.9$, $44.4$&$79.6$, $-28.1$&&$61 - 113\tablenotemark{g}$ & $13.7$ &$187$ & \nodata & \nodata &&$18.57$ &$187$ &$40$ & \nodata &&$22.9-30.9$  
\enddata
\tablenotetext{a}{Magellanic Stream coordinate system, defined in \citealt{Nidever2008}.}
\tablenotetext{b}{\ha\ intensities (not extinction-corrected). These are reported as $\iha\pm 1\sigma^{+\sigma_{\rm FS}}_{-\sigma_{\rm FS}}$, where $\sigma_{\rm FS}$ is the spread accounting for fits degenerate in $\widetilde{\chi}_{\rm min}^2$ due to an insufficient velocity range of the observations. This allows us to accurately anchor the continuum fit (see Section \ref{section:analysis}).}
\tablenotetext{c}{The extinction correction  for I$_{H\alpha}$ defined as $f_{\rm ext,~corr}=(e^{\rm A(\rm H\alpha)/2.5} - 1)\times100$, where $A(\rm H\alpha)$ is the total extinction.} 
\tablenotetext{d}{LAB \hi\ Survey smoothed to a $1\arcdeg$ angular resolution to match the WHAM observations.} 
\tablenotetext{e}{For the \hi\ emission, this assumes $\rm FWHM = 3\sigma \times \rm FWHM_{\rm H\textsc{~i}}$, where $\sigma$ is the standard deviation of the continuum. For the \ha\ emission, this assumes ${\rm FWHM}_{\rm H\alpha}=\rm FWHM_{\rm H\textsc{~i}}$ when \hi\ emission is detected or ${\rm FWHM}_{\rm H\alpha}=30~\kms$ when it is not detected.} 
\tablenotetext{f}{The PBV03(1) \& (2) sightlines probe the compact HVCs CHVC+266.0-18.7+336 and HVC+310.5+44.2+187, respectively. The \ha\ and \hi\ line properties were taken directly from \citet{Putman2003}.  Using the \citet{BH1999, BH2001} model of the Galactic ionizing radiation field ($f_{esc} \approx \rm 6\%$ normal to the disk), they find PBV03(1) to be at $1.2\le d_{\odot}\le 6.1~\kpc$ and PBV03(2) to be at $0.4\le d_{\odot}\le27.5~\kpc$.} 
\tablenotetext{g}{The \ha\ intensity is contaminated by an atmospheric line. \citet{Putman2003} estimate the error to be between $15$ and $30~{\rm mR}$.} 
\end{deluxetable*}

\section{Observations and Reduction}\label{obs}
\subsection{Sample}
In the spring of 2014, we observed the LA~Ext in \ha\ along $7$~sightlines  with the Wisconsin \ha~ Mapper (WHAM) at the Cerro-Tololo Inter-American Observatory in La Serena, Chile. The positions of our observations, sightlines~(a) through~(g), are shown in the \hi\ map in Figure~\ref{fig:all}. Our \ha\ observations span a Galactic longitude ($l$), Galactic latitude ($b$), and LSR velocity range of $(l,~b,~\vlsr) = (234\fdg1,$ $33\fdg5,$ $+50~\kms)$ to $(248\fdg9,$ $ 51\fdg6,$ $+250~\kms)$. In Figure~\ref{fig:all} and Table~\ref{table:intensities}, we also give the positions of our observations in the Magellanic Stream coordinate system, which has $l_{\rm MS} = 0\dg$ at the center of the LMC and the  $b_{\rm MS} = 0\dg$ line bisecting the Stream \citep{Nidever2008}. The Magellanic Stream coordinate system is useful because it eliminates the strong distortions of the MSys in standard equatorial and Galactic coordinates. In this coordinate system, our observations span $(l_{MS},~b_{MS}) = (76\fdg3,$ $11\fdg6$) to $(89\fdg3,$ $30\fdg5$). In this region, we specifically targeted six \hi\ cloudlets and one location off the \hi\ emission (see Figure~\ref{fig:all}). Three of these observations were intentionally aligned with UV bright background QSOs that were used in the \citet{Fox2014,Fox2018} UV absorption-line studies that explored the physical and chemical properties of this gas. These include sightlines~(c), (d), and~(e), which overlap with their HE1003+0149, IRASF09539-0439, LBQS1019+0147 sightlines, respectively.

\subsection{Observations}
The \ha\ emission from the LA is extremely faint and has previously been detected along only $2$~sightlines (see Table~\ref{table:intensities}), but even those detections were not along the main complexes, LA~I-IV (see Figure~\ref{fig:all}). The WHAM instrument was designed to observe faint ($I_{\rm H\alpha}\ge 0.03~{\rm R}$\footnote{$1~{\rm Rayleigh} = 10^{6} / {4 \pi}~ \textrm{photons} \cm^{-2} \sr^{-1} \s^{-1}$, which corresponds to $I_{\rm H\alpha}=2.41\times10^{-7}\,{\rm erg} \cm^{-2}\s^{-1}\sr^{-1}$. To convert \ha\ intensities to emission measures, $1~{\rm R}=2.75~(T_e/{10^4~{\rm K}})^{0.924}~\cm^{6}~{\rm pc}$, where $T_e$ is the electron temperature \citep{Barger2017}.}) optical emission from diffuse gas in the Galactic disk and halo. With its dual-etalon Fabry-P\'{e}rot optics, combined with a $0.6~{\rm m}$ objective lens, WHAM achieves a 1\dg~ angular resolution and $12~\kms$ velocity resolution ($\rm R \approx 25,000$) over a $200~\kms$ window near \ha\ (\citealt{Tufte1997, Reynolds1998, Haffner2003}).

\begin{figure*}
\begin{center}
\includegraphics[scale=0.5,angle=0]{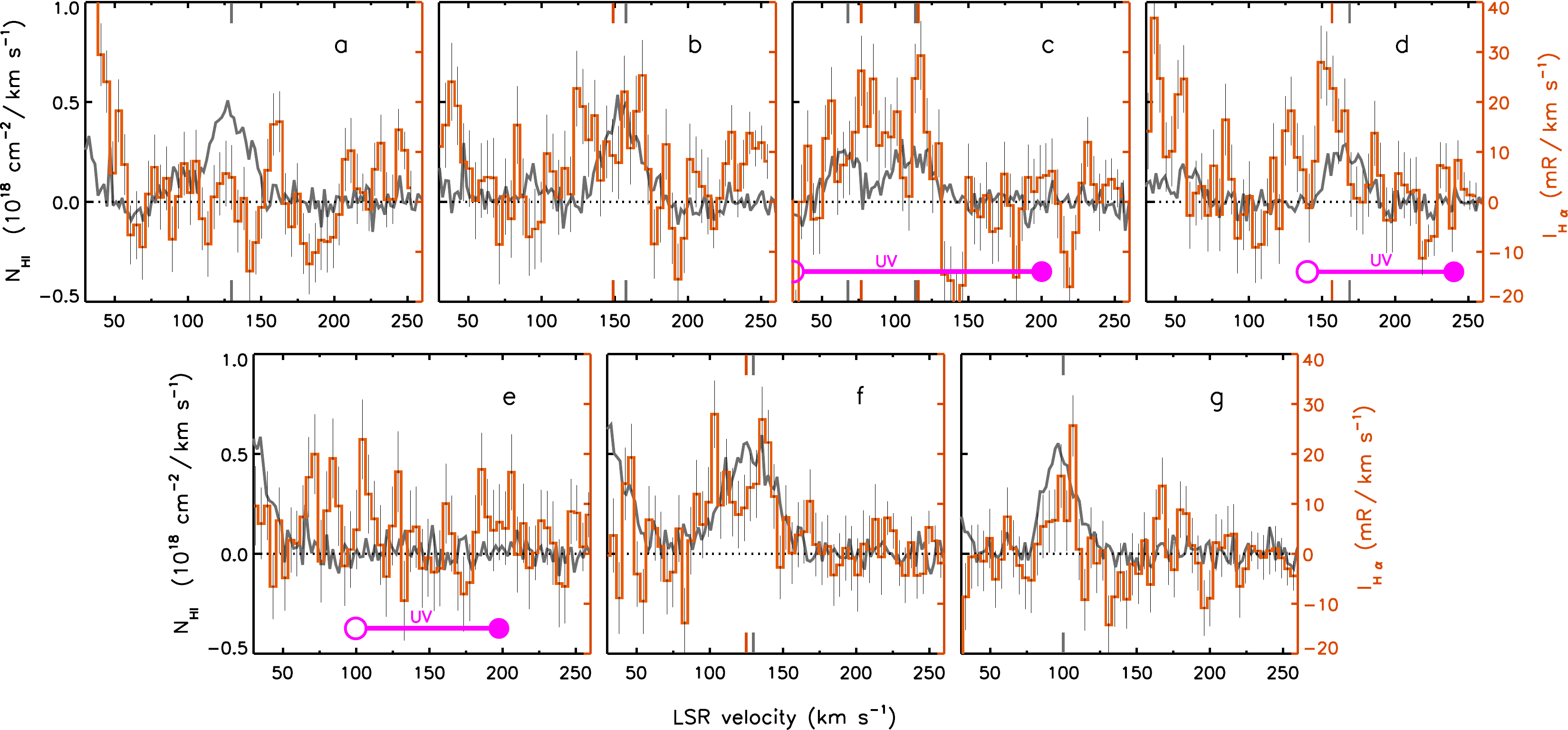}
\end{center}
\figcaption{\hi\ and \ha\ spectra toward our sightlines plotted in gray and orange, respectively. Color-coded line markers at the top and bottom of each subplot denote the velocity centroid of each type of emission. \hi, \ha, and UV components are all detected only toward sightline~(d). \ion{Si}{2}, \ion{Si}{3}, and \ion{Si}{4} absorption occurs in the $+140~\lesssim\vlsr\lesssim+240~\kms$ range, and this is shown by the pink marker. The open circle represents an upper limit on the smaller velocity as the \ion{Si}{2} and \ion{Si}{3} lines in (d) are blended with MW absorption and saturated (see Figure 7 in \citealt{Fox2014}). \label{figure:spec}}
\end{figure*}

We observed the LA~Ext over the course of three nights in 2014. On April~4th, we observed over a fixed geocentric (GEO) velocity range of $+100\le {\rm v_{GEO}}\le+300~\kms$, corresponding to $+65\lesssim\vlsr\lesssim+265~\kms$. On May~26th and~27th, we observed over a fixed $+60\le {\rm v_{GEO}}\le+260~\kms$ velocity range, corresponding to $+20\lesssim\vlsr\lesssim+220~\kms$. The combined observations span $+20\le\vlsr\le+265~\kms$. In Figure~\ref{figure:spec}, we only include the emission above $\vlsr\ge+35~\kms$ as the MW's emission is prominent at lower velocities.

Each of our sightlines was observed for a total integrated exposure time of $15$ -- $19~{\rm minutes}$. The ``on-target'' and the paired ``off-target'' observations were each taken over a 60-second duration such that one ``on-off'' pair consisted of 120 seconds of exposure time. The off-target observations were all positioned within $12\arcdeg$ of the on-target observations, at least $5\arcdeg$ from the $\log(\rm H\textsc{~i}/\cm^{-2})=18.7$ emission, and at least $0.55\arcdeg$ away from bright foreground stars ($m_{\rm V} < 6~{\rm mag}$). We subtracted the off-target observations from the on-target ones to remove the atmospheric contribution from our spectra. 

\subsubsection{Data Reduction, Velocity Calibration, and Extinction Correction}\label{subsection:reduction}
We reduced our \ha\ data with the standard WHAM pipeline for bias subtraction, flat-fielding, ring-summing, and cosmic ray contamination removal (see \citealt{Haffner2003} for more details). Our reduced observations have an overall \ha\ sensitivity of $I_{\rm H\alpha}\approx 30~{\rm mR}$, assuming a $30~\kms$ line width as measured in the simplest, resolved intermediate-velocity cloud (IVC) and HVC \ha\ profiles (e.g., \citealt{Tufte1998, Putman2003, Barger2012, Barger2017}).

We used the velocity calibration and atmospheric line subtraction process described in \citet{Barger2017} and the extinction correction procedure described in \citet{Barger2013,Barger2017}, which is briefly summarized below. The Leading Arm's complex velocity field is well mapped out along its \hi~21-cm emission (e.g., \citealt{Nidever2008}).
Using this as a guide, we centered the velocity window of our observations on the \hi\ emission by ``tuning'' the pressure of the ${\rm SF}_6$ gas between the etalons, since the linear relationship between SF6 pressure and wavelength has been well calibrated for WHAM \citep{Tufte1997}. 

 The geocentric velocities of the gas were calibrated by converting the monitored etalon gas pressures to wavelengths using this linear relationship and then measuring their Doppler shifts relative to $\lambda_{\rm H\alpha}=6562.8~\mbox{\AA}$. The calibration technique we employed is essentially the reverse of the tuning process that is described in \citet{Haffner2003} and is the same velocity calibration technique that was employed by \citet{Barger2017} for WHAM \ha\ observations along the Magellanic Stream. We used this technique because our observations did not cover a wide enough velocity window to include the peak of a bright atmospheric line positioned at a known geocentric velocity, which includes the OH atmospheric line centered at $\vgeo=+272.44~\kms$. Velocities calibrated in this study are accurate to ${\rm v}_{\rm GEO}\lesssim 5~\kms$ \citep{Madsen2004} as they were all taken with the same tune.\footnote{WHAM observations that are taken with different pressures, but the same interference order, will agree within ${\rm v}_{\rm radial}\lesssim 0.1~\kms$ of each other \citep{Barger2017}.} We account for this systematic uncertainty by adding $5~\kms$ in quadrature to the uncertainties of our line positions, i.e., $\sigma_{\rm total,~\rm H\alpha}=\sqrt{\sigma_{\rm fit}^2+(5~\kms)^2}$. These calibrated geocentric velocities were converted to the LSR frame by adding a constant velocity offset value. 

We corrected our \ha\ intensities for dust extinction associated with the Galactic interstellar dust along our sightlines, but not for self extinction. Because the LA~Ext gas is diffuse, accounting for self extinction will not change our \hi\ column density estimates appreciably. \citet{Barger2013} found an average extinction of 1.5\%  for the Magellanic Bridge (see their Table 2), where the average \hi\ column density is an order of magnitude higher than that of the LA Ext. By extension, self extinction would increase our \ha\ intensities by $\leq$ 1 mR, making its effect negligible. Using a MW dust law absorption-to-reddening ratio of the diffuse ISM of $R_{\rm v} = 3.1$ \citep{Cardelli1989}, we assume that the dust extinction is proportional to the \hi\ emission: 
\begin{equation}
A(H\alpha) = 5.14 \times 10^{-22}~ \langle  N_{\rm H\textsc{~i}}\rangle~\cm^{-2} ~{\rm atoms}^{-1}~{\rm mag} 
\end{equation}
\noindent where \avN\ is the spatially averaged column density of Galactic \hi\ along our line of sight, integrated over $-150 \lesssim\vlsr\lesssim+150~\kms$. The corrected \ha\ intensity is then
\begin{equation}
\iha_{\rm ,~corr} = \iha_{\rm ,~obs} e^{\frac{A(H\alpha)}{2.5}}
\end{equation}
On average, the extinction correction for the LA~Ext observations is 6.7\%, with a standard deviation of 1.4\% (see Table~\ref{table:intensities}). Such small extinction is expected, since all the sightlines are at $b \geq 30\dg$ above the Galactic equator. We apply this correction to our \ha\ intensities prior to our analysis in Sections~\ref{section:compare} and~\ref{section:dist}. The uncertainty in the extinction, determined by calculating \iha\ with $\widetilde{\chi}^{2} \leq 1.1\widetilde{\chi}^{2}_{\rm min}$ for each sightline (Section \ref{section:analysis}), is $0.5\%$. We do not propagate this uncertainty when determining the $I_{\rm H\alpha} \rightarrow \phi_{\rm LyC}$ distances (Section \ref{section:dist}), because it is small compared to the uncertainties from the spectral fitting and the radiation model. \\

\section{Data Analysis}\label{section:analysis}
To explore the properties of the neutral and ionized gas phases in the LA~Ext, we compare our \ha\ observations with archival \hi~21-cm observations from the  Leiden/Argentine/Bonn (LAB) survey \citep{Kalberla2005} and the UV absorption line results from \citet{Fox2018}. Although higher resolution data from the Effelsberg-Bonn \hi~ Survey (EBHIS) is available \citep{Kerp2016}, LAB's $0.6\dg$ beam size is a closer match to WHAM's $1\dg$ beam. 
The \ha\ and \hi\ spectra for our seven targeted observations, sightlines~(a) through~(g), are shown in Figure~\ref{figure:spec}. 

The LA~Ext \ha\ and \hi\ emission were modeled as a Gaussian that was combined with a linear fit to describe the continuum. For the \ha\ line fit, the Gaussian was also convolved with the WHAM instrument profile (see \citealt{Haffner2003}). For the \hi\ emission, we converted brightness temperatures into column densities using the following relationship\footnote{This relationship holds for self-absorption in diffuse \hi\ clouds that have an optical depth that is less than~$1$ (i.e., $\int \tau_{\rm v,~{\rm H\textsc{~i}}}~d{\rm v}<1$).}:
\begin{equation}\label{eq:nhi}
\mnhi=1.82\times10^{18 }~{\rm cm}^{-2} \int_{\rm v_{min}}^{\rm v_{max}} T_{B}({\rm v})~d{\rm v} 
\end{equation}
We fit the \hi\ and \ha\ data using the IDL MPFIT package \citep{Markwardt2009}, which incorporates the Levenberg-Marquardt non-linear least squares algorithm to minimize the reduced chi squared ($\widetilde{\chi}^2$)of the fit. For all sightlines, we assumed a single Gaussian fit unless $\widetilde{\chi}^2_{\rm min}$ exceeded 2.5 --- an indication that the data was under-fitted --- or if the \hi\ spectrum revealed complex substructure consisting of more than one component, as in the case of sightline~(c). 

For the \ha\ spectra, there was often insufficient continuum surrounding the emission line to perform the fit. This led to degeneracies in fit solutions with $\widetilde{\chi}^2\approx\widetilde{\chi}_{\rm min}^2$. To account for this, we use the procedure outlined in \citet{Barger2017}: We define a second uncertainty, the ``fitting spread ($\sigma_{FS}$),'' as the difference between the minimum and maximum \iha\ for models with $\widetilde{\chi}^2$ within $\rm 10\%$ of $\widetilde{\chi}^{2}_{\rm min}$. Hence, the \ha\ intensities are reported as $\iha\pm 1\sigma^{+\sigma_{\rm FS}}_{-\sigma_{\rm FS}}$ in Table~\ref{table:intensities}. This spread was determined by doing a grid search for the continuum fit and Gaussian parameters. For instance, sightline~(f) has a best fit  \ha\ intensity of $45\pm3~{\rm mR}$ with $\widetilde{\chi}^{2}_{\rm min}$ = 1.0 and a range of intensities that satisfy $\widetilde{\chi}^{2} \leq~ 1.1~\widetilde{\chi}^{2}_{min} $ between $30\lesssim I_{H\alpha} \lesssim 59~{\rm mR}$. 

We report \ha\ and \hi\ emission only when we detect $I_{\rm H\alpha}~{\rm or}~N_{\rm H\textsc{~i}}\geq 3\sigma\times {\rm FWHM}$ --- where $\sigma$ here is the standard deviation of the continuum. For the \hi\ emission, we assumed a representative line width of ${\rm FWHM}=30~\kms$ and required that the \ha\ emission has a width that is at least as wide as the \hi\ emission or $30~\kms$ when \hi\ emission is not detected (as in sightline (e)). Table~\ref{table:intensities} summarizes the best fit \hi\ column densities, \ha\ intensities, and their corresponding line widths and velocity centroids.

\section{Line Strengths and Kinematics in Neutral and Ionized Phases}\label{section:compare}

Multi-wavelength data in the radio, optical, and UV toward the same sightline enables us to explore the multi-phased nature of this gas. In this section, we first compare the emission strengths and kinematics of our \ha\ observations with those of archival \hi\ observations. These line properties are listed in Table~\ref{table:intensities}. We then compare the \hi\ and \ha\ emission with the UV metal-line absorption along sightlines~(c), (d), and~(e).

There are a number of properties worth noting about the cloudlets we observe. Some of  these properties are in line with what is expected for classical Galactic HVCs, and others agree better with the line properties of disturbed systems like the Magellanic Bridge and Stream. First, from the FWHM values, we find that the \ha\ emission is up to twice as broad as the 21-cm emission. Although in general, ionized gas line widths are broader than those for neutral gas, that is usually not the case in classical HVCs (e.g., Complex M, C, A, \citealt{Tufte1998, Barger2012}). The only known exception to this is Complex K \citep{Haffner2001}. Conversely, in more complex systems like the Magellanic Bridge, the \ha\ line widths are noticeably larger \citep{Barger2012}. This may indicate that the \ha\ emission is tracing gas that is more turbulent (see Section 2.2 in \cite{Barger2012} for a more detailed discussion of the different scenarios under which this is possible. Second, there is a significant detection of \hi\ emission toward sightlines~(a), and~(g), but no \ha\ emission detected at the $3\sigma$ confidence level, although  \ha\ is detected at $2\sigma$ toward sightline~(g). Third, there is no correlation between the estimated \ha\ intensities and \hi\ column densities, although the line centroids agree to within $13~\kms$. This property of our observed emission is generally in agreement with what is expected for HVCs, since  HVCs tend to exhibit  uncorrelated \ha\ and \hi\ emission strengths (e.g., \citealt{Tufte1998, Haffner2001,Haffner2005}, \citetalias{Putman2003}, \citealt{Hill2009, Barger2012, Barger2013, Barger2017}).  

\citet{Fox2014} studied the UV absorption toward background QSOs in three of our sightlines. They reported no LA absorption along sightlines~(c) and~(e), which overlap with QSOs HE1003+0149 and LBQS1019+0147, respectively. Their study focuses on the $+200\lesssim\vlsr\lesssim+300~\kms$ portion of the spectra along those two particular sightlines, whereas in this study we have centered our observations at $\vlsr\approx+150~\kms$ based on the \hi\ Gaussian decompositions of \citet{Nidever2008}. \citet{Fox2014} focused on this higher velocity range because the absorption of the low ionization species (e.g., C\textsc{~ii}, S\textsc{~ii}, and Si\textsc{~ii}) is saturated and too blended with the MW's absorption to easily be distinguished. However, along sightline~(c), there is absorption at $+40\lesssim\vlsr\le+200~\kms$, where there are two \hi\ emission lines present. Similarly, there is absorption present in these low ionization species along sightline~(e) at  $+100\lesssim\vlsr\le+200~\kms$ (see Figure~\ref{figure:spec}). The asymmetric shape of the blended absorption from $-60\lesssim\vlsr\lesssim+180~\kms$ along sightline~(c) and from $-80\lesssim\vlsr\lesssim+200~\kms$ along sightline~(e) (Sightlines HE1003+0149 and LBQS1019+0147 in Figure 7 of \citealt{Fox2014}) strongly suggests that, in addition to MW absorption, there is also absorption that is related to a MW intermediate velocity cloud (IVC), HVC, and/or the LA~Ext. The asymmetric shape of the absorption in the  high-ionization species Si\textsc{~iv} and C\textsc{~iv} along sightline~(c) suggests that this cloudlet is highly ionized. Along sightline~(e), there are also weak hints of absorption in these high ionization species. 

The UV absorption along sightline~(d) is also blended with the MW and hence, difficult to separate, though \citet{Fox2014,Fox2018} indicated that the LA was detected in absorption along this sightline. The C\textsc{~ii}, Si\textsc{~ii}, Si\textsc{~iii}, and Si\textsc{~iv} absorption spans $+140\lesssim\vlsr \leq +240~\kms$, where the lower velocity limit of the LA contribution is estimated from the wing of an \hi\ emission line centered at $\vlsr=+166\pm1~\kms$. We mark the kinematic extent of the absorption toward background quasar IRASF09539-0439  in Figure~\ref{figure:spec}. This absorption overlaps with both the \hi\ and \ha\ emission. The detection of high ionization species toward sightlines~(c) and (d) in addition to both \hi\ and \ha\ indicates that this cloud is multi-phased and is made up of gas with an electron temperature of $10^{3.9}\lesssim T_e \lesssim 10^{5.3}~{\rm K}$ (e.g., \citealt{Gnat2007, Oppenheimer2013}). 
\section{Distance to the Leading Arm Extension and other LA HVC\lowercase{s}}\label{section:dist}

Because \ha\ is a recombination line, its line strength is directly proportional to the rate of ionization per surface area of the emitting gas in photoionization equilibrium conditions. Using this premise, we place constraints on the distance to the LA~Ext using an equation derived in \citet{Barger2012} that relates the \ha\ intensity of a gas cloud with the source of its ionization:
\begin{equation}\label{eq:FHalpha}
\phi_{\rm H\textsc{~i}\rightarrow H\textsc{~ii}} = 2.1\times10^5\left(\frac{\iha}{0.1~{\rm R}}\right)\left(\frac{T_e}{10^4~\rm K}\right)^{0.118} {\rm photons}\ {\rm cm}^{-2}\ {\rm s}^{-1}.
\end{equation}
We assume that the incident ionizing flux is dominated by photons from OB stars which escape the MW's ISM and that the LA~Ext cloudlets are optically thick to the Lyman continuum and optically thin to \ha\ photons (since $\log{\left(N_{\rm H\textsc{~i}}/\cm^{-2}\right)}\ge 18$). With this assumption, $\phi_{\rm H\textsc{~i}\rightarrow H\textsc{~ii}}\approx\phi_{\rm{LyC}}$, where $\phi_{\rm{LyC}}$ is the incident Lyman Continuum flux.  We also include the small --- but not negligible --- ionizing contributions from the MCs and the low-redshift ($z \approx 0$) extragalactic background (EGB).

We use the \citet{Fox2014} model for the combined Galactic, Magellanic, and extragalactic ionizing radiation field. This model is an updated version of the \citet{Fox2005} model, which is based on those of \citet{BH1999, BH2001, BH2002} and \citet{Barger2013}. Here, the fraction of hard photons escaping normal to the disk is $f_{\rm esc,~MW} \approx 6\%$. For the ionizing contribution of the LMC and SMC, $f_{\rm esc,~LMC} = 3~\pm~1.0\%$ and $f_{\rm esc,~SMC} = 4~\pm~1.5\%$. 

The MW, LMC, and SMC ionization models specifically predict the amount of radiation that is escaping out of the disks of these galaxies into the circumgalactic medium. To estimate how far the ionizing radiation extends from the host galaxy disk, these flux models were calibrated using distances to HVCs and IVCs obtained via the direct methods described in Section \ref{section:introduction}. For the Milky Way, this calibration was done using \ha\ emission-line observations of Complexes~A, M, C, and K and distance measurements from absorption-line spectroscopy (see \citealt{BH1999, BH2002, BH2004}). This calibration enables one to use make reasonable estimates of HVC distances using the $I_{\rm H\alpha} \rightarrow \phi_{\rm Lyc}$ distance method (e.g., \citealt{Putman2003, Barger2012,Barger2017}). The escape fraction for the Milky Way has a factor of 2 uncertainty, which could change the model fluxes by up to 50\% \citet{BH2002}. Although these uncertainties are large, the Milky Way model has produced distances to the Smith Cloud \citep{Putman2003} and Complex A \citep{Barger2012} that are in agreement with kinematic distance estimates \citep{Lockman2008} and stellar absorption distance brackets \citep{Ryans1997, Wakker1996,Wakker2008, VW1999}.

The LMC and SMC flux models were calibrated using \ha\ observations of the Magellanic Bridge and the well-measured distances of the LMC and SMC \citep{Barger2013}. We include the ionizing contribution of the EGB, which is estimated to be $\phi_{\rm{LyC,~EGB}}\lesssim 10^{4}~\rm photons~cm^{-2}~s^{-1}$ at $z=0$ \citep{Weymann2001}, which will raise the \iha\ by up to $5~{\rm mR}$ using Equation~(\ref{eq:FHalpha}).  We assume that the EGB is isotropic, and given that the cloudlets are optically thin to \ha\ and LyC photons, that the combined model is independent of observer axis (that is, it is adequate for determining the combined radiation of the MW, MCs, and EGB incident on an HVC).

\begin{figure}
\begin{center}
\includegraphics[trim={0.9cm 0.7cm 0.7cm 0.8cm},clip ,scale=0.32,angle=0]{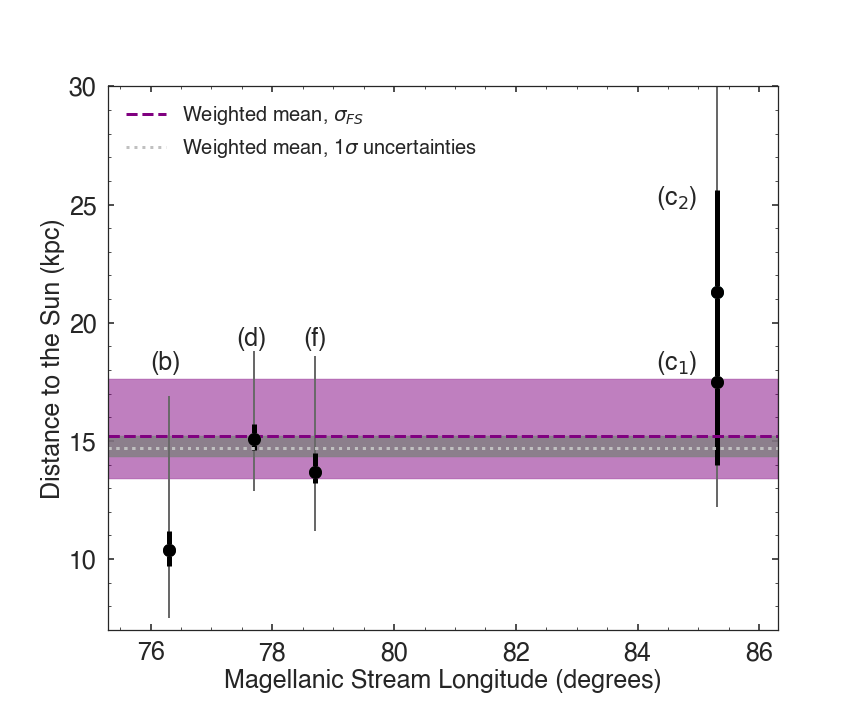}
\end{center}
\figcaption{Heliocentric distances to the LA~Ext cloudlets with \ha~ detections as a function of Magellanic Stream longitude. We mark the statistical uncertainties with thick lines and the spread in distances from fits that have $\widetilde{\chi}^2\le 1.1~\widetilde{\chi}_{\rm min}^2$ ($\sigma_{FS}$) with thin lines. The gray and purple bands represent the uncertainties on the weighted mean distances due to the $\rm 1\sigma$ statistical uncertainties and spread in fitted intensities, respectively. \label{fig:dist}}
\end{figure}

\begin{figure*}
\begin{center}
\includegraphics[scale=0.35]{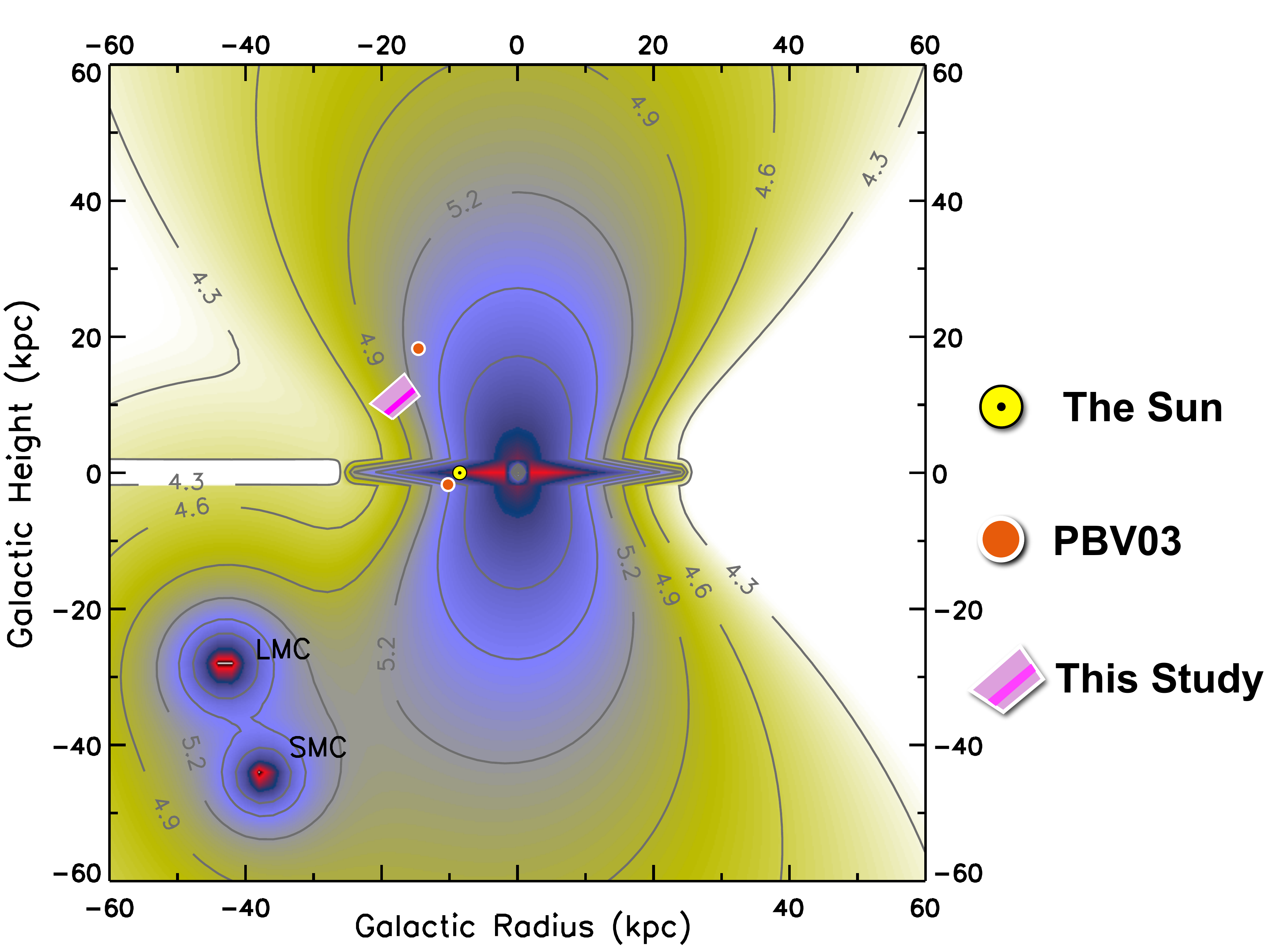}
\caption{A $120\times120~\kpc$ slice through the three-dimensional Lyman continuum flux model centered on the Milky Way. This radiation field includes contributions from the Milky Way \citep{Fox2005}, Magellanic Clouds \citep{Barger2013}, and extragalactic background at $z \approx 0$ \citep{Weymann2001}. The contour lines are $\log{\left(\phi_{\rm LyC}\right)}$ in $\rm photons~cm^{-2}~ s^{-1}$. The locations in the Galactic halo at which  \ha\ emission was detected towards the LA Ext and the \cite{Putman2003} HVCs --- CHVC+266.0-18.7+336 and HVC+310.5+44.2+187 --- are shown as the pink polygon and orange dots, respectively. The Sun is marked as a yellow dot for reference. Because there could be other sources of ionization besides photoionization, we consider the distances derived using this method as lower distance limits.
\label{fig:contour}}
\end{center}
\end{figure*}

We determined the ionizing flux needed to reproduce the observed \ha\ intensities using Equation~(\ref{eq:FHalpha}) and then used the combined model of the ionizing radiation from the MW, MCs, and EGB (Figure \ref{fig:contour}) to determine the distances of the cloudlets from the Sun. However, since there are other sources of ionization that we may not have accounted for, this approach only provides a lower distance limit. That is, if there are other sources of ionization contributing to the observed \ha\ emission, the contribution from photoionization would be less, which would place the clouds at a larger distance. Other possible types of ionization occuring in the clouds include collisional ionization associated with ram-pressure stripping and self-ionization from a ``shock cascade'' created when parts of the cloud that have been stripped and decelerated by the halo collide with the other parts of the cloud  (see \citealt{BH2007, TG2015,Barger2017}).

We quantify the effects of two sources of uncertainty on the distance to each sightline: the statistical uncertainty and the spread in intensities ($\sigma_{FS}$) associated with degenerate fits, which have $\widetilde{\chi}^2\ \approx \widetilde{\chi}_{\rm min}^2$ (see Section~\ref{section:analysis}). We calculated the uncertainties on the LA Ext distance due to the $1\sigma$ statistical uncertainties, as follows: 
\begin{equation}
\begin{split}
\sigma^{+} &=   d({\iha + 1\sigma}) - d({\iha})\\ \notag
\sigma^{-} &= d({\iha}) - d({\iha-1\sigma})\notag
\end{split}
\end{equation}
The uncertainties on the distance due to $\sigma_{FS}$ were calculated in the same fashion. For this, we consider all fits satisfying $\widetilde{\chi}_{\rm min}^2\le \widetilde{\chi}^2\le 1.1~\widetilde{\chi}_{\rm min}^2$. Therefore, each sightline has a distance given by $d ^{1\sigma^{+}~(+\sigma_{FS})}_{1\sigma^{-}~(-\sigma_{FS})}$. For example, sightline (f) is at $d_{\odot}=13.7^{+0.8}_{-0.5}~\kpc$ for $\iha=41\pm4~{\rm mR}$, and its range of distances when accounting for the spread in intensities for fits degenerate in $\widetilde{\chi}^2$ is $11.2\le d_{\odot}\le18.4~\kpc$ for $31\le\iha\le52~{\rm mR}$ (see Table~\ref{table:intensities}). Calculating the spread in intensities for each sightline also  allows us to take into consideration variation in the \ha\ intensities of up to 50\%. This way, the uncertainties in the radiation model are accounted for in our distance estimates.

Since the size of the explored LA~Ext region is small compared to the size of the entire LA, we assume that all of the LA~Ext cloudlets are at the same distance in order to calculate a weighted mean distance. To account for the asymmetry of the uncertainties in distance, we use the technique described in \citet{Barlow2003}.\footnote{We used the \citet{Barlow2003} Java applet found at \url{http://www.slac.stanford.edu/\%7Ebarlow/java/statistics5.html}} We find a weighted distance of $d_{\odot}= 14.7^{+0.5}_{ -0.4}~\kpc$ using the $1\sigma$ statistical uncertainties and $d_{\odot}=15.2^{ +2.4}_{ -1.8}~\kpc$ using the spread in intensities from multiple fits degenerate in $\tilde{\chi}^2$. We show the \ha\ distances for each sightline as well as the resultant weighted distances as a function of Magellanic Stream longitude in Figure~\ref{fig:dist}. Because the method we employed can only be used to place a lower limit on the distance to these clouds, {\em the LA Extension is at least $13.4~\kpc$ from the Sun}.\footnote{Excluding the contribution from the MCs changes the distance of the LA~Ext from $d_\odot \ge 13.4~\kpc$ to $d_\odot \ge 16.7~\kpc$.}

In determining this lower distance limit, we treated the \ha\ components along sightline~(c) as separate sightlines, as shown in Figure~\ref{fig:dist}.  It is uncertain if both are actually physically  associated with the LA~Ext. The lower velocity \hi\ component of sightline~(c) at $\vlsr=+73\pm7~\kms$ (denoted c$_1$ in Figure \ref{fig:dist}) is kinematically offset from the \ha\ and \hi\ emission features along our other sightlines by $\Delta\vlsr\approx-40~\kms$. This deviation from the bulk motion of the LA Ext may indicate that it is a Galactic~IVC. Excluding this lower velocity component of sightline~(c) from our calculations increases $d_{\odot}$ by $0.43~\kpc$.

Additionally, it is important to note that the LA Ext overlaps with Wannier complex WB, which covers $(l,~b,~\vlsr)=(225\dg,~0\dg,~+70~\kms)$ to $(265\dg,~60\dg,~+170~\kms)$ (\citealt{Wannier1972a, Wakker1991}). Using absorption-line spectroscopy toward a background star, \citet{Thom2006} placed the distance of this cloud at $d_{\odot}\lesssim 8.8 ^{+2.3}_{-1.3}~\kpc$. Therefore, based on our \ha\ distances, the LA~Ext cloudlets explored in this study are unlikely to be associated with Wannier Complex~WB.

As mentioned in Section~\ref{section:introduction}, prior to this study, there were only two \ha\ detections in gas clouds that might be associated with the LA. These detections, which were made by \cite{Putman2003}, are listed in Table~\ref{table:intensities}. Using Fabry-P{\'e}rot \ha\ observations from the Anglo-Australian Telescope and the \cite{BH1999,BH2001} model for the ionizing flux from OB stars in the disk of the Milky Way, \cite{Putman2003} derived distances to CHVC+266.0-18.7+336 and HVC+310.5+44.2+187. The former is located in the same part of the sky covered by LA~IV and the latter on the outskirts of LA~II (see Figure~\ref{fig:all}). We used the combined MW, MCs, and EGB flux models for the incident ionizing radiation and the \cite{Putman2003} \ha\ intensities to rederive the distances to these HVCs (Table~\ref{table:intensities}). We include a summary of all $\rm H\alpha$ distances against a slice through the radiation model in Figure~\ref{fig:contour}.  

 \section{Discussion}\label{section:disc}

\begin{figure*}
\begin{center}
\includegraphics[trim={1.9cm 8.3cm 1.5cm 5cm},clip, scale=0.55]{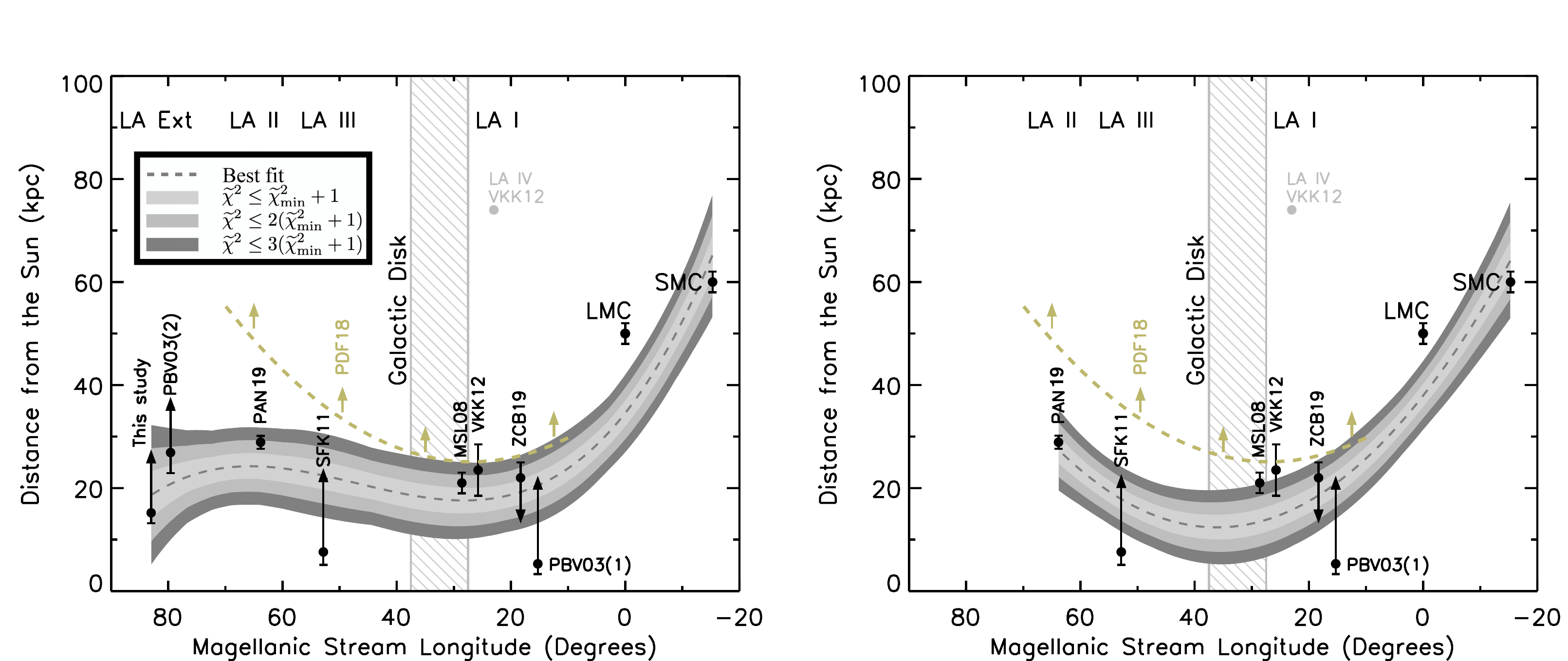}
\end{center}
\figcaption{A compilation of all LA distance constraints and their 1$\sigma$ uncertainties. The literature distances and their corresponding abbreviations are as follows: \cite{Putman2003} (\citetalias{Putman2003}), \cite{PriceWhelan2019} (\citetalias{PriceWhelan2019}), \cite{Smoker2011} (\citetalias{Smoker2011}), \cite{Mcclure-Griffiths2008} (\citetalias{Mcclure-Griffiths2008}), \cite{Venzmer2012}(\citetalias{Venzmer2012}), and \cite{Zhang2019} (\citetalias{Zhang2019}). The dashed gray line marks our best fit distance, obtained by randomly sampling the parameter space covered by the distances and fitting polynomials of various orders to the draws (Section \ref{section:disc}). We fixed the maximum allowable distance for each upper limit at 50 kpc, based on what simulations predict for an HVC {\em leading} the Magellanic Clouds. The light gray envelope illustrates fits that are within ${\widetilde{\chi}}^2\le \widetilde{\chi}_{\rm min}^2+1$, and the progressively darker envelopes are for fits with ${\widetilde{\chi}}^2$ values ~$2$ and~$3$ times that of the light gray envelope. We also include the line-of-sight distances predicted by the \cite{Pardy2018} simulation (abbreviated \citetalias{Pardy2018}) as dashed olive green line, but we do not use them to constrain our models. {\bf Left:} Best fit  model including our LA Ext and HVC+310.5+44.2+187 distances. {\bf Right:} Best fit model excluding the aforementioned, as they may have a non-Magellanic origin. In both cases, simulations predict higher line-of-sight distances than observed.
\label{fig:mlon_dsun}}
\end{figure*} 
 
The Leading Arm is patchy and covers one-eighth of the sky ($\sim 5600$ square degrees). The fragmented nature of the LA and its large size present challenges to finding distances along its entire length. Nevertheless, a few studies have been able to place constraints on the distance to LA I \citep{Mcclure-Griffiths2008, Putman2003, Venzmer2012, Zhang2019} and LA II \citep{Smoker2011, PriceWhelan2019} using radio and optical spectroscopy, as well as optical imaging. The details and limitations of these methods are discussed in Section \ref{section:introduction}. All LA distance measurements in the literature, with the exception of those from \cite{Mcclure-Griffiths2008}, \cite{PriceWhelan2019}, and the \cite{Venzmer2012} LA I estimate are either upper or lower limits due to limitations of the methods used. The distances inferred from the \cite{Putman2003} \ha\ measurements and our LA Ext \ha\ measurements are lower distance limits for reasons discussed in Section~\ref{section:dist}. Similarly, the distance inferred from the UV absorption observed toward background star CD14-A05 \citep{Fox2018} at $d_{\odot}=22 \pm 3~\kpc$ (\cite{Zhang2019}), provides a $\rm 1\sigma$ upper limit of $d_{\odot}<25~\kpc$ to the absorbing gas. We have plotted all of these LA distance measurements and their 1$\sigma$ uncertainties as a function of Magellanic Stream longitude in Figure~\ref{fig:mlon_dsun}. 

Based on predictions from the extensive suite of simulations of the  dynamical history of the Magellanic Clouds (see \citealt{DOnghia2016} for a review), we can assume that because LA~I, II and~III are leading the orbital path of the two galaxies, and because the Magellanic Clouds are just past perigalacticon in their orbits, these complexes and all of their associated cloud fragments are expected to be at a distance closer than that of the Clouds. That is, the LA observations should be at a distance of $d\lesssim55~\kpc$ from the Galactic Center. \cite{Venzmer2012} estimated a distance of $d_{\odot}\approx 74~\kpc$ to LA~IV. Based on the reasoning above, however, this distance is inconsistent with what is expected for an HVC complex that is {\em leading} the Magellanic Clouds. Furthermore, although there are \hi\ studies that have suggested a kinematic association of LA~IV (\citealt{Venzmer2012}, \citealt{For2013, For2016}) with the LA, no one has yet investigated the chemical abundance patterns in that region to confirm membership. This is mainly because searches for $ z < 1$ QSOs towards LA~IV clouds along sightlines with \hi\ emission have been unsuccessful. For these reasons, we do not include the \cite{Venzmer2012} distance to LA~IV in Figure~\ref{fig:mlon_dsun}.

In simulations of the Magellanic System (MSys), predicted heliocentric distances are often compared with measured distances as a function of longitude to judge the quality of the model(s) used . Because the LA has fewer distance estimates than the other parts of the MSys, this comparison was for a long time done using only the \cite{Mcclure-Griffiths2008} distance (\citealt{Besla2012}, \citealt{Diaz2011}, \citealt{Guglielmo2014}, \citealt{Pardy2018}), and more recently, the \cite{PriceWhelan2019} distance \citep{TepperGarca2019}. With the compilation of all the LA distances in the literature, in addition to our three new ones, we make this comparison along the entire length of the LA for the first time. To determine whether or not the observed trend in distance to the LA matches predictions from simulations, we produced a fit to the measured distances (Figure \ref{fig:mlon_dsun}). Because we did not want to make any assumptions about the functional form of the two-dimensional structure of the LA, we fit polynomials of various orders using MPFIT to random draws from the parameter space covered by the measurements. This way, we sampled all possible fits in the distance ranges covered by the measurements and their corresponding limits and 3$\sigma$ uncertainties. The maximum allowed distance for the LA was set to $50~\kpc$, based on the orbit of the Magellanic Clouds. We would like to point out that the fit should be viewed more as an approximation and a visual aid than an exact solution. This is because we have not modeled and accounted for the non-Gaussianity in the upper and lower limits. Additionally, we are more interested in the general distance trend suggested by the fit. For this reason, we also show fits with up to $3(\widetilde{\chi}^{2}_{min} + 1$) in Figure \ref{fig:mlon_dsun}. Readers interested producing a model with a more rigorous treatment of the upper and lower limits can refer to the methodology of \citep{Isobe1986}. 

One thing that is evident from Figure \ref{fig:mlon_dsun} is that the distance to the Leading Arm varies from its base near the Clouds to its tip in the halo. Using the excellent spatial resolution offered by the GASS survey, \citet{For2016} reported the first observational evidence of this variation in distance. The compilation of distances to the LA in Figure \ref{fig:mlon_dsun} not only appears to lend support to this picture, but also provides constraints which can be used to estimate the ages of the LA HVCs at those locations. Combined, ages and distances can provide strong constraints on the {\em origin of the progenitor clouds of the Leading Arm.}

The exact origin of the LA is still a subject of debate. Although the fit in Figure \ref{fig:mlon_dsun} was not fixed to either of the Magellanic Clouds, the best fit line passes through the SMC and not the LMC. This is an interesting result because it is consistent with the findings of \citet{For2016}, \citet{Fox2018}, and \citet{Richter2018}. Recent simulations suggest that the LA contains material from both the LMC and the SMC \citep{Pardy2018}. While they are able to broadly reproduce the observed velocities and positions of the HVCs belonging to the LA, they predict higher heliocentric distances than observed. Because of this, and because the LA exhibits a large spread in metallicities and ages \citep{Fox2018}, some have suggested that the LA may have multiple origins. \citet{DOnghia2008} proposed that the Magellanic Clouds could be the two largest galaxies in an accreted group of dwarfs. They estimated that seven of the eleven brightest MW satellites may have been part of this accreted group. More recently, \citet{Koposov2015a, Koposov2015b, Koposov2018} and \citet{DrlicaWagner2016} have discovered ten ultra-faint galaxies in proximity to the Magellanic Clouds. In this scenario, the LA could be gaseous debris from material that was stripped from dwarf galaxies that are in front of the orbital path of the Magellanic Clouds \citep{Yang2014, Hammer2015}. This LA formation scenario is consistent with the variation of the chemical composition pattern that \citet{Fox2018} and \citet{Richter2018} found for LA~I, II, III, and~ the LA Ext, and could explain why the LA lacks an extended and old tidal stellar stream. Furthermore, the only simulations that are able to reproduce LA~IV are those that invoke the dwarf galaxy scenario. If it is true that some parts of the LA originated from gas-bearing dwarfs, this might explain why LA~IV is extremely fragmented and why it might lie at a much higher distance than other LA complexes ($d_{\odot}\approx 74~\kpc$; \citealt{Venzmer2012}).

Although the metallicity of the LA Ext has been shown to be consistent with a Magellanic origin, \citet{TepperGarca2019} point out that the original chemical signature of the LA cloudlets could be washed out due to mixing. Additionally, they show that it is highly unlikely for an HVC to survive the ram pressure stripping due to the hot Galactic halo. They suggest based on this, that LA-North ($l_{MS} \lesssim 65\arcdeg$) would need to be protected in a deep dark matter potential, such as a dwarf galaxy, to survive those conditions. On the contrary, others have demonstrated that it is possible for a neutral hydrogen gas cloud to survive the conditions in the Galactic halo for several Myr provided it is large ($> 250\; \rm pc$), although it will end up fragmented \citep{Armillotta2017}.

If the the LA Ext and HVC+310.5+44.2+187 are part of the Leading Arm, then the line of best fit places the closest point of approach of the LA at $d_{\odot}\approx20~\kpc$, and it occurs at $l_{\rm MS}=+80\arcdeg$. If they are not, then the minimum distance is $d_{\odot}\approx15~\kpc$, and it occurs at $l_{\rm MS}=+36\arcdeg$. The position of the point of closest approach in the right panel of Figure \ref{fig:mlon_dsun} is consistent with predictions from \cite{Pardy2018}, however their minimum distance is higher. Figure \ref{fig:mlon_dsun} also demonstrates there is still much uncertainty in the distances to the Leading Arm. Many of the distance measurements are derived through indirect methods, and some are only able to provide upper or lower distance bounds. We stress that more direct distance measurements of the LA are needed to constrain the three-dimensional structure of the Leading Arm, its total mass, and the rate at which it is accreting onto the Milky Way. Nevertheless, our compilation of Leading Arm distances to date provides important constraints that will inform future simulations of the dynamical history and orbital motion of the Magellanic Clouds. 


\section{Summary}
In this study, we have presented Wisconsin \ha~ Mapper observations toward seven cloudlets in the LA Extension (LA~Ext), a region recently found by \citet{Fox2018} to be chemically consistent with the Leading Arm. Toward~$4$ out of~$7$ of our sightlines, we detected faint \ha\ emission with intensities $33\leq \iha \leq~49~{\rm mR}$ at velocities $+73\leq\vlsr\leq+154~\kms$ (Table~\ref{table:intensities}). Three of these sightlines align with UV-bright background QSOs that were used by \citet{Fox2014, Fox2018} to explore this gas through absorption-line spectroscopy. The detection of \hi\ and \ha\ in emission as well as low and high ionization species in absorption (C\textsc{~ii}, Si\textsc{~ii}, Si\textsc{~iii}, Si\textsc{~iv}, and C\textsc{~iv}) indicate that the LA~Ext gas is multi-phased. 

We estimated the distance to the LA~Ext by assuming that photoionization is the dominant ionizing mechanism and that the incident ionizing radiation field includes contributions from the Milky Way, Magellanic Clouds, and extragalactic background at z$\sim$0. Using this radiation model (Figure \ref{fig:contour}), we determined that the LA~Ext is at a heliocentric distance of $d_{\odot}\ge13.4~\kpc$ from the Sun. This distance places it at a height of $|z| \gtrsim 8.6~\kpc$ above the Galactic disk. We also rederived the \ha\ distances in \cite{Putman2003} for CHVC+266.0-18.7+336 and HVC+310.5+44.2+187 with this model and found that they are located at $d_{\odot}\ge 5.0~\kpc$ and $d_{\odot} \ge 22.9~\kpc$, respectively. If these two clouds are associated with the LA, then the above represent distances to LA~IV and the gas on the outskirts of LA~II. 

Finally, we combined our estimates with distance measurements in the literature to show how the distance to the LA varies along its length (Figure~\ref{fig:mlon_dsun}). If HVC+310.5+44.2+187 and the LA~Ext are indeed associated with the LA, then the minimum LA distance is $d_{\odot}\approx20~\kpc$ and occurs at $l_{\rm MS}=+80\arcdeg$. 
\setlength\baselineskip{11pt} 
\acknowledgements{We thank the anonymous referee for their constructive report which greatly helped to improve the quality of this paper. We are also grateful to Andrew Fox for feedback on an earlier draft, and to Josh Peek for useful conversations on the origin of the stars in LA~I. WHAM operations for these observations were supported by National Science Foundation (NSF) awards AST 1108911 and AST 1714472/1715623 and AST 1940634. This paper includes archived LAB and GASS \hi\ data obtained through the AIfA \hi\ Surveys Data Server (https://www.astro.uni-bonn.de/hisurvey/index.php). The IDL routines are available at \url{http://purl.com/net/mpfit}. Jacqueline Antwi-Danso received additional support through NSF grant PHY-1358770 and Kat Barger through NSF Astronomy and Astrophysical Postdoctoral Fellowship award AST~1203059.}

\software{MPFIT \citep{Markwardt2009}}
\facility{WHAM (University of Wisconsin, Madison Wisconsin H-Alpha Mapper)}


\begin{thebibliography}{}
\expandafter\ifx\csname natexlab\endcsname\relax\def\natexlab#1{#1}\fi
\providecommand{\url}[1]{\href{#1}{#1}}
\providecommand{\dodoi}[1]{doi:~\href{http://doi.org/#1}{\nolinkurl{#1}}}
\providecommand{\doeprint}[1]{\href{http://ascl.net/#1}{\nolinkurl{http://ascl.net/#1}}}
\providecommand{\doarXiv}[1]{\href{https://arxiv.org/abs/#1}{\nolinkurl{https://arxiv.org/abs/#1}}}

\bibitem[{{Armillotta} {et~al.}(2017){Armillotta}, {Fraternali}, {Werk},
  {Prochaska}, \& {Marinacci}}]{Armillotta2017}
{Armillotta}, L., {Fraternali}, F., {Werk}, J.~K., {Prochaska}, J.~X., \&
  {Marinacci}, F. 2017, \mnras, 470, 114, \dodoi{10.1093/mnras/stx1239}

\bibitem[{{Barger} {et~al.}(2013){Barger}, {Haffner}, \&
  {Bland-Hawthorn}}]{Barger2013}
{Barger}, K.~A., {Haffner}, L.~M., \& {Bland-Hawthorn}, J. 2013, \apj, 771,
  132, \dodoi{10.1088/0004-637X/771/2/132}

\bibitem[{{Barger} {et~al.}(2012){Barger}, {Haffner}, {Wakker}, {Hill},
  {Madsen}, \& {Duncan}}]{Barger2012}
{Barger}, K.~A., {Haffner}, L.~M., {Wakker}, B.~P., {et~al.} 2012, \apj, 761,
  145, \dodoi{10.1088/0004-637X/761/2/145}

\bibitem[{{Barger} {et~al.}(2017){Barger}, {Madsen}, {Fox}, {Wakker},
  {Bland-Hawthorn}, {Nidever}, {Haffner}, {Antwi-Danso}, {Hernandez}, {Lehner},
  {Hill}, {Curzons}, \& {Tepper-Garc{\'{\i}}a}}]{Barger2017}
{Barger}, K.~A., {Madsen}, G.~J., {Fox}, A.~J., {et~al.} 2017, \apj, 851, 110,
  \dodoi{10.3847/1538-4357/aa992a}

\bibitem[{Barlow(2003)}]{Barlow2003}
Barlow, R. 2003, PhyStat2003, 250

\bibitem[{{Besla} {et~al.}(2007){Besla}, {Kallivayalil}, {Hernquist},
  {Robertson}, {Cox}, {van der Marel}, \& {Alcock}}]{Besla2007}
{Besla}, G., {Kallivayalil}, N., {Hernquist}, L., {et~al.} 2007, \apj, 668,
  949, \dodoi{10.1086/521385}

\bibitem[{Besla {et~al.}(2012)Besla, Kallivayalil, Hernquist, {Van Der Marel},
  Cox, \& Kere{\v{s}}}]{Besla2012a}
Besla, G., Kallivayalil, N., Hernquist, L., {et~al.} 2012, Mon. Not. R. Astron.
  Soc, 421, 2109, \dodoi{10.1111/j.1365-2966.2012.20466.x}

\bibitem[{{Besla} {et~al.}(2010){Besla}, {Kallivayalil}, {Hernquist}, {van der
  Marel}, {Cox}, \& {Kere{\v s}}}]{Besla2010}
{Besla}, G., {Kallivayalil}, N., {Hernquist}, L., {et~al.} 2010, \apjl, 721,
  L97, \dodoi{10.1088/2041-8205/721/2/L97}

\bibitem[{{Besla} {et~al.}(2012){Besla}, {Kallivayalil}, {Hernquist}, {van der
  Marel}, {Cox}, \& {Kere{\v{s}}}}]{Besla2012}
---. 2012, \mnras, 421, 2109, \dodoi{10.1111/j.1365-2966.2012.20466.x}

\bibitem[{{Bland-Hawthorn} \& {Maloney}(1999)}]{BH1999}
{Bland-Hawthorn}, J., \& {Maloney}, P.~R. 1999, \apjl, 510, L33,
  \dodoi{10.1086/311797}

\bibitem[{{Bland-Hawthorn} \& {Maloney}(2001)}]{BH2001}
---. 2001, \apjl, 550, L231, \dodoi{10.1086/319654}

\bibitem[{{Bland-Hawthorn} \& {Maloney}(2002)}]{BH2002}
---. 2002, Astronomical Society of the Pacific Conference Series, Vol. 254,
  {H{\ensuremath{\alpha}} Distance Constraints for High Velocity Clouds in the
  Galactic Halo}, 267

\bibitem[{{Bland-Hawthorn} \& {Putman}(2004)}]{BH2004}
{Bland-Hawthorn}, J., \& {Putman}, M.~E. 2004, in IAU Symposium, Vol. 217,
  Recycling Intergalactic and Interstellar Matter, ed. P.-A. {Duc},
  J.~{Braine}, \& E.~{Brinks}, 12

\bibitem[{{Bland-Hawthorn} {et~al.}(2007){Bland-Hawthorn}, {Sutherland},
  {Agertz}, \& {Moore}}]{BH2007}
{Bland-Hawthorn}, J., {Sutherland}, R., {Agertz}, O., \& {Moore}, B. 2007,
  \apjl, 670, L109, \dodoi{10.1086/524657}

\bibitem[{{Br{\"u}ns} {et~al.}(2005){Br{\"u}ns}, {Kerp}, {Staveley-Smith},
  {Mebold}, {Putman}, {Haynes}, {Kalberla}, {Muller}, \&
  {Filipovic}}]{Bruns2005}
{Br{\"u}ns}, C., {Kerp}, J., {Staveley-Smith}, L., {et~al.} 2005, \aap, 432,
  45, \dodoi{10.1051/0004-6361:20040321}

\bibitem[{Cardelli {et~al.}(1989)Cardelli, Clayton, \& Mathis}]{Cardelli1989}
Cardelli, J.~A., Clayton, G.~C., \& Mathis, J.~S. 1989, \apj, 345, 244

\bibitem[{Casetti-Dinescu {et~al.}(2014)Casetti-Dinescu, Bidin, Girard,
  M{\'{e}}ndez, Vieira, Korchagin, \& van Altena}]{Casetti-Dinescu2014}
Casetti-Dinescu, D.~I., Bidin, C.~M., Girard, T.~M., {et~al.} 2014, \apj, 784,
  L37, \dodoi{10.1088/2041-8205/784/2/L37}

\bibitem[{Chiappini(2008)}]{Chiappini2008}
Chiappini, C. 2008, in ASP Conference Series, Vol. 396

\bibitem[{{Diaz} \& {Bekki}(2011)}]{Diaz2011}
{Diaz}, J., \& {Bekki}, K. 2011, \mnras, 413, 2015,
  \dodoi{10.1111/j.1365-2966.2011.18289.x}

\bibitem[{D'Onghia \& Fox(2016)}]{DOnghia2016}
D'Onghia, E., \& Fox, A.~J. 2016, \aapr, 54, 363,
  \dodoi{10.1146/annurev-astro-081915-023251}

\bibitem[{{D'Onghia} \& {Lake}(2008)}]{DOnghia2008}
{D'Onghia}, E., \& {Lake}, G. 2008, \apjl, 686, L61, \dodoi{10.1086/592995}

\bibitem[{{Drlica-Wagner} {et~al.}(2016){Drlica-Wagner}, {Bechtol}, {Allam},
  {Tucker}, {Gruendl}, {Johnson}, {Walker}, {James}, {Nidever}, {Olsen},
  {Wechsler}, {Cioni}, {Conn}, {Kuehn}, {Li}, {Mao}, {Martin}, {Neilsen},
  {Noel}, {Pieres}, {Simon}, {Stringfellow}, {van der Marel}, \&
  {Yanny}}]{DrlicaWagner2016}
{Drlica-Wagner}, A., {Bechtol}, K., {Allam}, S., {et~al.} 2016, \apjl, 833, L5,
  \dodoi{10.3847/2041-8205/833/1/L5}

\bibitem[{For {et~al.}(2013)For, Staveley-Smith, \&
  Mcclure-Griffiths}]{For2013}
For, B.-Q., Staveley-Smith, L., \& Mcclure-Griffiths, N.~M. 2013, \apj, 764

\bibitem[{For {et~al.}(2016)For, Staveley-Smith, Mcclure-Griffiths, Westmeier,
  \& Bekki}]{For2016}
For, B.-Q., Staveley-Smith, â.~L., Mcclure-Griffiths, â. N.~M., Westmeier,
  â.~T., \& Bekki, K. 2016, MNRAS, 461, 892, \dodoi{10.1093/mnras/stw1364}

\bibitem[{{Fox} {et~al.}(2005){Fox}, {Wakker}, {Savage}, {Tripp}, {Sembach}, \&
  {Bland-Hawthorn}}]{Fox2005}
{Fox}, A.~J., {Wakker}, B.~P., {Savage}, B.~D., {et~al.} 2005, \apj, 630, 332,
  \dodoi{10.1086/431915}

\bibitem[{{Fox} {et~al.}(2014){Fox}, {Wakker}, {Barger}, {Hernandez},
  {Richter}, {Lehner}, {Bland-Hawthorn}, {Charlton}, {Westmeier}, {Thom},
  {Tumlinson}, {Misawa}, {Howk}, {Haffner}, {Ely}, {Rodriguez-Hidalgo}, \&
  {Kumari}}]{Fox2014}
{Fox}, A.~J., {Wakker}, B.~P., {Barger}, K.~A., {et~al.} 2014, \apj, 787, 147,
  \dodoi{10.1088/0004-637X/787/2/147}

\bibitem[{{Fox} {et~al.}(2018){Fox}, {Barger}, {Wakker}, {Richter},
  {Antwi-Danso}, {Casetti-Dinescu}, {Howk}, {Lehner}, {D'Onghia}, {Crowther},
  \& {Lockman}}]{Fox2018}
{Fox}, A.~J., {Barger}, K.~A., {Wakker}, B.~P., {et~al.} 2018, \apj, 854, 142,
  \dodoi{10.3847/1538-4357/aaa9bb}

\bibitem[{{Gnat} \& {Sternberg}(2007)}]{Gnat2007}
{Gnat}, O., \& {Sternberg}, A. 2007, \apjs, 168, 213, \dodoi{10.1086/509786}

\bibitem[{{Guglielmo} {et~al.}(2014){Guglielmo}, {Lewis}, \&
  {Bland-Hawthorn}}]{Guglielmo2014}
{Guglielmo}, M., {Lewis}, G.~F., \& {Bland-Hawthorn}, J. 2014, \mnras, 444,
  1759, \dodoi{10.1093/mnras/stu1549}

\bibitem[{{Haffner}(2005)}]{Haffner2005}
{Haffner}, L.~M. 2005, in Astronomical Society of the Pacific Conference
  Series, Vol. 331, Extra-Planar Gas, ed. R.~{Braun}, 25

\bibitem[{{Haffner} {et~al.}(2001){Haffner}, {Reynolds}, \&
  {Tufte}}]{Haffner2001}
{Haffner}, L.~M., {Reynolds}, R.~J., \& {Tufte}, S.~L. 2001, \apjl, 556, L33,
  \dodoi{10.1086/322867}

\bibitem[{Haffner {et~al.}(2003)Haffner, Reynolds, Tufte, Madsen, Jaehnig, \&
  Percival}]{Haffner2003}
Haffner, L.~M., Reynolds, R.~J., Tufte, S.~L., {et~al.} 2003, \apj, 149, 405

\bibitem[{{Hammer} {et~al.}(2015){Hammer}, {Yang}, {Flores}, {Puech}, \&
  {Fouquet}}]{Hammer2015}
{Hammer}, F., {Yang}, Y.~B., {Flores}, H., {Puech}, M., \& {Fouquet}, S. 2015,
  \apj, 813, 110, \dodoi{10.1088/0004-637X/813/2/110}

\bibitem[{Hill {et~al.}(2009)Hill, Haffner, \& Reynolds}]{Hill2009}
Hill, A.~S., Haffner, L.~M., \& Reynolds, R.~J. 2009, \apj, 703, 1832,
  \dodoi{10.1088/0004-637X/703/2/1832}

\bibitem[{{Isobe} {et~al.}(1986){Isobe}, {Feigelson}, \& {Nelson}}]{Isobe1986}
{Isobe}, T., {Feigelson}, E.~D., \& {Nelson}, P.~I. 1986, \apj, 306, 490,
  \dodoi{10.1086/164359}

\bibitem[{Kalberla {et~al.}(2005)Kalberla, Burton, Hartmann, Arnal, Bajaja,
  Morras, \& Poppel}]{Kalberla2005}
Kalberla, P. M.~W., Burton, W.~B., Hartmann, D., {et~al.} 2005, \astap, Volume
  440, Issue 2, September III 2005, pp.775-782, 440, 775,
  \dodoi{10.1051/0004-6361:20041864}

\bibitem[{{Koposov} {et~al.}(2015{\natexlab{a}}){Koposov}, {Belokurov},
  {Torrealba}, \& {Evans}}]{Koposov2015a}
{Koposov}, S.~E., {Belokurov}, V., {Torrealba}, G., \& {Evans}, N.~W.
  2015{\natexlab{a}}, \apj, 805, 130, \dodoi{10.1088/0004-637X/805/2/130}

\bibitem[{{Koposov} {et~al.}(2015{\natexlab{b}}){Koposov}, {Casey},
  {Belokurov}, {Lewis}, {Gilmore}, {Worley}, {Hourihane}, {Randich}, {Bensby},
  {Bragaglia}, {Bergemann}, {Carraro}, {Costado}, {Flaccomio}, {Francois},
  {Heiter}, {Hill}, {Jofre}, {Lando}, {Lanzafame}, {de Laverny}, {Monaco},
  {Morbidelli}, {Sbordone}, {Mikolaitis}, \& {Ryde}}]{Koposov2015b}
{Koposov}, S.~E., {Casey}, A.~R., {Belokurov}, V., {et~al.} 2015{\natexlab{b}},
  \apj, 811, 62, \dodoi{10.1088/0004-637X/811/1/62}

\bibitem[{{Koposov} {et~al.}(2018){Koposov}, {Walker}, {Belokurov}, {Casey},
  {Geringer-Sameth}, {Mackey}, {Da Costa}, {Erkal}, {Jethwa}, {Mateo},
  {Olszewski}, \& {Bailey}}]{Koposov2018}
{Koposov}, S.~E., {Walker}, M.~G., {Belokurov}, V., {et~al.} 2018, \mnras, 479,
  5343, \dodoi{10.1093/mnras/sty1772}

\bibitem[{Lockman {et~al.}(2008)Lockman, Benjamin, Heroux, \&
  Langston}]{Lockman2008}
Lockman, F.~J., Benjamin, R.~A., Heroux, A.~J., \& Langston, G.~I. 2008, \apjl,
  Volume 679, Issue 1, article id. L21, pp. (2008)., 679,
  \dodoi{10.1086/588838}

\bibitem[{{Lu} {et~al.}(1994){Lu}, {Savage}, \&
  {Sembach}}]{1994ApJ...426..563L}
{Lu}, L., {Savage}, B.~D., \& {Sembach}, K.~R. 1994, \apj, 426, 563,
  \dodoi{10.1086/174093}

\bibitem[{{Lu} {et~al.}(1998){Lu}, {Savage}, {Sembach}, {Wakker}, {Sargent}, \&
  {Oosterloo}}]{1998AJ....115..162L}
{Lu}, L., {Savage}, B.~D., {Sembach}, K.~R., {et~al.} 1998, \aj, 115, 162,
  \dodoi{10.1086/300181}

\bibitem[{{Madsen}(2004)}]{Madsen2004}
{Madsen}, G.~J. 2004, PhD thesis, The University of Wisconsin - Madison,
  Wisconsin, USA

\bibitem[{{Markwardt}(2009)}]{Markwardt2009}
{Markwardt}, C.~B. 2009, in Astronomical Society of the Pacific Conference
  Series, Vol. 411, Astronomical Society of the Pacific Conference Series, ed.
  {D.~A.~Bohlender, D.~Durand, \& P.~Dowler}, 251

\bibitem[{McClure-Griffiths {et~al.}(2010)McClure-Griffiths, Madsen, Gaensler,
  McConnell, \& Schnitzeler}]{McClure-Griffiths2010}
McClure-Griffiths, N.~M., Madsen, G.~J., Gaensler, B.~M., McConnell, D., \&
  Schnitzeler, D. H. F.~M. 2010, \apj, 725, 275,
  \dodoi{10.1088/0004-637X/725/1/275}

\bibitem[{Mcclure-Griffiths {et~al.}(2008)Mcclure-Griffiths, Staveley-Smith,
  Lockman, Calabretta, Ford, Kalberla, Murphy, Nakanishi, \&
  Pisano}]{Mcclure-Griffiths2008}
Mcclure-Griffiths, N.~M., Staveley-Smith, L., Lockman, F.~J., {et~al.} 2008,
  \apj, 673, 143

\bibitem[{McClure-Griffiths {et~al.}(2009)McClure-Griffiths, Pisano,
  Calabretta, Ford, Lockman, Staveley-Smith, Kalberla, Bailin, Dedes,
  Janowiecki, Gibson, Murphy, Nakanishi, \&
  Newton-McGee}]{McClure-Griffiths2009}
McClure-Griffiths, N.~M., Pisano, D.~J., Calabretta, M.~R., {et~al.} 2009,
  \apjs, Volume 181, Issue 2, pp. 398-412 (2009)., 181, 398,
  \dodoi{10.1088/0067-0049/181/2/398}

\bibitem[{Nidever {et~al.}(2008)Nidever, Majewski, \& Burton}]{Nidever2008}
Nidever, D.~L., Majewski, S.~R., \& Burton, W.~B. 2008, \apj, 679, 432,
  \dodoi{10.1086/587042}

\bibitem[{Nidever {et~al.}(2010)Nidever, Majewski, Burton, \&
  Nigra}]{Nidever2010}
Nidever, D.~L., Majewski, S.~R., Burton, W.~B., \& Nigra, L. 2010, \apj, 723,
  1618, \dodoi{10.1088/0004-637X/723/2/1618}

\bibitem[{Oppenheimer \& Schaye(2013)}]{Oppenheimer2013}
Oppenheimer, B.~D., \& Schaye, J. 2013, MNRAS, 434, 1043,
  \dodoi{10.1093/mnras/stt1043}

\bibitem[{{Pardy} {et~al.}(2018){Pardy}, {D'Onghia}, \& {Fox}}]{Pardy2018}
{Pardy}, S.~A., {D'Onghia}, E., \& {Fox}, A.~J. 2018, \apj, 857, 101,
  \dodoi{10.3847/1538-4357/aab95b}

\bibitem[{{Price-Whelan} {et~al.}(2019){Price-Whelan}, {Nidever}, {Choi},
  {Schlafly}, {Morton}, {Koposov}, \& {Belokurov}}]{PriceWhelan2019}
{Price-Whelan}, A.~M., {Nidever}, D.~L., {Choi}, Y., {et~al.} 2019, \apj, 887,
  19, \dodoi{10.3847/1538-4357/ab4bdd}

\bibitem[{Putman {et~al.}(2003)Putman, Bland-Hawthorn, Veilleux, Gibson,
  Freeman, \& Maloney}]{Putman2003}
Putman, M.~E., Bland-Hawthorn, J., Veilleux, S., {et~al.} 2003, \apj, 597, 948

\bibitem[{Putman {et~al.}(1998)Putman, Gibson, Staveley-Smith, Banks, Barnes,
  Bhatalk, Disney, Ekers, Freeman, Haynes, Henning, Jerjen, Kilborn,
  Koribalski, Knezek, Malin, Mould, Oosterloo, Price, Ryder, Sadler, Stewart,
  Stootmank, Vailek, Webster, \& Wright}]{Putman1998}
Putman, M.~E., Gibson, B.~K., Staveley-Smith, L., {et~al.} 1998, Nature, 394

\bibitem[{Putman {et~al.}(2002)Putman, {De Heij}, Staveley-Smith, Braun,
  Freeman, Gibson, Burton, Barnes, Banks, Bhathal, {De Blok}, Boyce, Disney,
  Drinkwater, Ekers, Henning, Jerjen, Kilborn, Knezek, Koribalski, Malin,
  Marquarding, Minchin, Mould, Oosterloo, Price, Ryder, Sadler, Stewart,
  Stootman, Webster, \& Wright}]{Putman2002}
Putman, M.~E., {De Heij}, V., Staveley-Smith, L., {et~al.} 2002, The
  Astronomical Journal, 123, 873

\bibitem[{Reynolds {et~al.}(1998)Reynolds, Tufte, Haffner, Jaehnig, \&
  Percival}]{Reynolds1998}
Reynolds, R.~J., Tufte, S.~L., Haffner, L.~M., Jaehnig, K., \& Percival, J.~W.
  1998, Publ. Astron. Soc. Aust, 15, 14

\bibitem[{{Richter} {et~al.}(2018){Richter}, {Fox}, {Wakker}, {Howk}, {Lehner},
  {Barger}, {D{'}Onghia}, \& {Lockman}}]{Richter2018}
{Richter}, P., {Fox}, A.~J., {Wakker}, B.~P., {et~al.} 2018, \apj, 865, 145,
  \dodoi{10.3847/1538-4357/aadd0f}

\bibitem[{{Richter} {et~al.}(2017){Richter}, {Nuza}, {Fox}, {Wakker}, {Lehner},
  {Ben Bekhti}, {Fechner}, {Wendt}, {Howk}, {Muzahid}, {Ganguly}, \&
  {Charlton}}]{Richter2017a}
{Richter}, P., {Nuza}, S.~E., {Fox}, A.~J., {et~al.} 2017, \aap, 607, A48,
  \dodoi{10.1051/0004-6361/201630081}

\bibitem[{{Ryans} {et~al.}(1997){Ryans}, {Keenan}, {Sembach}, \&
  {Davies}}]{Ryans1997}
{Ryans}, R.~S.~I., {Keenan}, F.~P., {Sembach}, K.~R., \& {Davies}, R.~D. 1997,
  \mnras, 289, 986, \dodoi{10.1093/mnras/289.4.986}

\bibitem[{{Sembach} {et~al.}(2001){Sembach}, {Howk}, {Savage}, \&
  {Shull}}]{2001AJ....121..992S}
{Sembach}, K.~R., {Howk}, J.~C., {Savage}, B.~D., \& {Shull}, J.~M. 2001, \aj,
  121, 992, \dodoi{10.1086/318777}

\bibitem[{{Smoker} {et~al.}(2011){Smoker}, {Fox}, \& {Keenan}}]{Smoker2011}
{Smoker}, J.~V., {Fox}, A.~J., \& {Keenan}, F.~P. 2011, \mnras, 415, 1105,
  \dodoi{10.1111/j.1365-2966.2011.18647.x}

\bibitem[{{Tepper-Garc{\'\i}a} {et~al.}(2019){Tepper-Garc{\'\i}a},
  {Bland-Hawthorn}, {Pawlowski}, \& {Fritz}}]{TepperGarca2019}
{Tepper-Garc{\'\i}a}, T., {Bland-Hawthorn}, J., {Pawlowski}, M.~S., \& {Fritz},
  T.~K. 2019, \mnras, 488, 918, \dodoi{10.1093/mnras/stz1659}

\bibitem[{{Tepper-Garc{\'{\i}}a} {et~al.}(2015){Tepper-Garc{\'{\i}}a},
  {Bland-Hawthorn}, \& {Sutherland}}]{TG2015}
{Tepper-Garc{\'{\i}}a}, T., {Bland-Hawthorn}, J., \& {Sutherland}, R.~S. 2015,
  \apj, 813, 94, \dodoi{10.1088/0004-637X/813/2/94}

\bibitem[{Thom {et~al.}(2006)Thom, Putman, Gibson, Christlieb, Flynn, Beers,
  Wilhelm, \& Lee}]{Thom2006}
Thom, C., Putman, M.~E., Gibson, B.~K., {et~al.} 2006, \apj, 638, 97

\bibitem[{{Tufte}(1997)}]{Tufte1997}
{Tufte}, S.~L. 1997, PhD thesis, THE UNIVERSITY OF WISCONSIN - MADISON

\bibitem[{{Tufte} {et~al.}(1998){Tufte}, {Reynolds}, \& {Haffner}}]{Tufte1998}
{Tufte}, S.~L., {Reynolds}, R.~J., \& {Haffner}, L.~M. 1998, \apj, 504, 773,
  \dodoi{10.1086/306103}

\bibitem[{{van Woerden} {et~al.}(1999){van Woerden}, {Schwarz}, {Peletier},
  {Wakker}, \& {Kalberla}}]{VW1999}
{van Woerden}, H., {Schwarz}, U.~J., {Peletier}, R.~F., {Wakker}, B.~P., \&
  {Kalberla}, P. M.~W. 1999, \nat, 400, 138, \dodoi{10.1038/22061}

\bibitem[{Venzmer {et~al.}(2012)Venzmer, Kerp, \& Kalberla}]{Venzmer2012}
Venzmer, M.~S., Kerp, J., \& Kalberla, P. M.~W. 2012, Astronomy {\&}
  Astrophysics, 547, A12, \dodoi{10.1051/0004-6361/201118677}

\bibitem[{Wakker {et~al.}(1996)Wakker, Howk, Schwarz, van Woerden, Beers,
  Wilhelm, Kalberla, \& Danly}]{Wakker1996}
Wakker, B., Howk, C., Schwarz, U., {et~al.} 1996, The Astrophysical Journal,
  473, 834

\bibitem[{Wakker \& van Woerden(1991)}]{Wakker1991}
Wakker, B., \& van Woerden, H. 1991, Astronomy {\&} Astrophysics, 250, 509

\bibitem[{{Wakker} {et~al.}(2008){Wakker}, {York}, {Wilhelm}, {Barentine},
  {Richter}, {Beers}, {Ivezi{\'c}}, \& {Howk}}]{Wakker2008}
{Wakker}, B.~P., {York}, D.~G., {Wilhelm}, R., {et~al.} 2008, \apj, 672, 298,
  \dodoi{10.1086/523845}

\bibitem[{Wannier {et~al.}(1972)Wannier, Wrixon, \& Wilson}]{Wannier1972a}
Wannier, P., Wrixon, G., \& Wilson, R. 1972, Astronomy {\&} Astrophysics, 18,
  224

\bibitem[{Weymann {et~al.}(2001)Weymann, Vogel, Veilleux, \&
  Epps}]{Weymann2001}
Weymann, R.~J., Vogel, S.~N., Veilleux, S., \& Epps, H.~W. 2001, \apj, 561, 559

\bibitem[{{Winkel} {et~al.}(2016){Winkel}, {Kerp}, {Fl{\"o}er}, {Kalberla},
  {Ben Bekhti}, {Keller}, \& {Lenz}}]{Kerp2016}
{Winkel}, B., {Kerp}, J., {Fl{\"o}er}, L., {et~al.} 2016, \aap, 585, A41,
  \dodoi{10.1051/0004-6361/201527007}

\bibitem[{{Yang} {et~al.}(2014){Yang}, {Hammer}, {Fouquet}, {Flores}, {Puech},
  {Pawlowski}, \& {Kroupa}}]{Yang2014}
{Yang}, Y., {Hammer}, F., {Fouquet}, S., {et~al.} 2014, \mnras, 442, 2419,
  \dodoi{10.1093/mnras/stu931}

\bibitem[{{Zhang} {et~al.}(2019){Zhang}, {Casetti-Dinescu}, {Moni Bidin},
  {M{\'e}ndez}, {Girard}, {Vieira}, {Korchagin}, {van Altena}, \&
  {Zhao}}]{Zhang2019}
{Zhang}, L., {Casetti-Dinescu}, D.~I., {Moni Bidin}, C., {et~al.} 2019, \apj,
  871, 99, \dodoi{10.3847/1538-4357/aaf560}

\end{thebibliography}

\end{document}